\documentclass[iop,onecolumn]{emulateapj}

\usepackage{natbib}
\usepackage{rotating}
\usepackage{epsfig}
\usepackage{psfig}
\usepackage{amsmath,amssymb}
\allowdisplaybreaks

\def\ArhoR{A_{\rho{R}}}
\newcommand\Arhoz{A_{\rho{z}}}
\newcommand\ApR{A_{pR}}
\newcommand\Apz{A_{pz}}
\newcommand\cmminusone{\,{\rm cm}^{-1}}
\newcommand\cmminusthree{\,{\rm cm}^{-3}}
\newcommand\czero{c_0}
\newcommand\czerotil{\tilde{c}_0}
\renewcommand\d{{\rm d}}
\newcommand\Dm{\mathcal{D}}
\newcommand\erg{\,{\rm erg}}
\newcommand\Fzero{{F_0}}

\newcommand\grad{{\nabla}}

\renewcommand\i{{\rm i}}
\newcommand\kms{{\rm \,km\,s^{-1}}}
\newcommand\kpc{{\rm \,kpc}}

\newcommand\kmin{k_{\rm min}}
\newcommand\kMRI{k_{\rm MRI}}
\newcommand\kz{k_z}
\newcommand\kR{k_R}
\newcommand\kztil{\tilde{k}_z}
\newcommand\kRtil{\tilde{k}_R}
\newcommand\ktil{\tilde{k}}
\newcommand\kphi{k_{\phi}}
\newcommand\kv{{\bf k}}
\newcommand\kvtil{\tilde{\bf k}}
\newcommand\kzksq{{\kz^2\over k^2}}

\newcommand\ntil{\tilde{n}}
\newcommand\ntilmax{\tilde{n}_{\rm max}}
\newcommand\nezero{n_{e,0}}
\newcommand\pd{\partial}
\newcommand\pzero{{p_0}}

\newcommand\sminusone{\,{\rm s}^{-1}}

\newcommand\Tzero{T_0}
\newcommand\vz{{v_z}}
\newcommand\vR{{v_R}}

\newcommand\vphi{v_{\phi}}
\newcommand\vRtil{{\tilde{v}_R}}
\newcommand\vztil{{\tilde{v}_z}}

\newcommand\vvzero{\vv_0}
\newcommand\vzerophi{v_{0\phi}}
\newcommand\vzeroR{v_{0R}}
\newcommand\vzeroz{v_{0z}}

\newcommand\vv{{\bf v}}

\newcommand\GammapR{\Gamma_{pR}}
\newcommand\Gammapz{\Gamma_{pz}}
\newcommand\GammarhoR{\Gamma_{\rho{R}}}
\newcommand\Gammarhoz{\Gamma_{\rho{z}}}
\newcommand\GammaOmegaR{\Gamma_{\Omega{R}}}
\newcommand\GammaOmegaz{\Gamma_{\Omega{z}}}
\newcommand\GammaTR{\Gamma_{TR}}
\newcommand\GammaTz{\Gamma_{Tz}}
\newcommand\GammaTp{\Gamma_{Tp}}

\newcommand\rhozero{\rho_0}
\newcommand\rhotil{\tilde{\rho}}

\newcommand\omegaczerotil{\tilde\omega_{\rm c,0}}
\newcommand\omegaA{\omega_{\rm A}}
\newcommand\omegaAsq{\omega^2_{\rm A}}

\newcommand\omegaAsqtil{\tilde{\omega}^2_{\rm A}}

\newcommand\omegadtil{\tilde{\omega}_{\rm d}}

\newcommand\omegaBVsq{\omega^2_{\rm BV}}
\newcommand\omegaBVsqtil{\tilde{\omega}^2_{\rm BV}}

\newcommand\omegarotsq{\omega^2_{\rm rot}}
\newcommand\omegacmag{\omega_{\rm c,mag}}
\newcommand\omegacmagtil{\tilde\omega_{\rm c,mag}}
\newcommand\omegacphisq{\omega^2_{\rm c,\phi}}
\newcommand\omegacphisqtil{\tilde\omega^2_{\rm c,\phi}}

\newcommand\omegarotsqtil{\tilde{\omega}^2_{\rm rot}}

\newcommand\omegac{\omega_{\rm c}}
\newcommand\omegactil{\tilde{\omega}_{\rm c}}

\newcommand\omegath{\omega_{\rm th}}
\newcommand\omegathtil{\tilde{\omega}_{\rm th}}

\newcommand\kelvin{\,{\rm K}} 

\renewcommand\bv{{\bf b}}
\newcommand\bR{b_{R}}
\newcommand\bz{b_{z}}
\newcommand\bphi{b_{\phi}}
\newcommand\de{\partial}

\newcommand\Bv{{\bf B}}
\newcommand\Bvzero{\Bv_0}
\newcommand\Bzero{B_0}
\newcommand\Bzerophi{B_{0\phi}}
\newcommand\BzeroR{B_{0R}}
\newcommand\Bzeroz{B_{0z}}
\newcommand\bzeroR{b_{0R}}
\newcommand\bzeroz{b_{0z}}
\newcommand\bzerophi{b_{0\phi}}
\newcommand\Bphi{B_{\phi}}
\newcommand\BR{B_{R}}
\newcommand\Bz{B_{z}}

\newcommand\Qv{{\bf Q}}
\newcommand\ds{\displaystyle}
\renewcommand\div{\nabla \cdot}
\newcommand\rot{\nabla \times}
\newcommand\e{{\rm e}}

\newcommand\bvzero{\bv_0}
\newcommand\vvA{\vv_{\rm A}}

\renewcommand\L{\mathcal{L}}
\newcommand\LT{\mathcal{L}_T}
\newcommand\Lrho{\mathcal{L}_{\rho}}
\newcommand\omegaca{\omega_{\rm c,a}}
\newcommand\omegacatil{\tilde\omega_{\rm c,a}}

\newcommand\Zsun{Z_{\odot}}
\newcommand\xcrit{x_{\rm crit}}

\newcommand\Xdot{\dot{X}}
\newcommand\XRR{X_{RR}}
\newcommand\XRz{X_{Rz}}
\newcommand\XRrho{X_{R\rho}}
\newcommand\XzR{X_{zR}}
\newcommand\Xzz{X_{zz}}
\newcommand\Xzrho{X_{z\rho}}
\newcommand\XrhoR{X_{\rho R}}
\newcommand\Xrhoz{X_{\rho z}}
\newcommand\Xrhorho{X_{\rho\rho}}
\newcommand\XRRdot{\dot{X}_{RR}}
\newcommand\XRzdot{\dot{X}_{Rz}}
\newcommand\XRrhodot{\dot{X}_{R\rho}}
\newcommand\XzRdot{\dot{X}_{zR}}
\newcommand\Xzzdot{\dot{X}_{zz}}
\newcommand\Xzrhodot{\dot{X}_{z\rho}}
\newcommand\YrhoR{Y_{\rho R}}
\newcommand\Yrhoz{Y_{\rho z}}
\newcommand{\idl}{{\sc idl }}
\newcommand{\fzroots}{{\sc fz\_roots}}

\defcitealias{Nip10}{N10}
\defcitealias{Nip13}{NP13}

\slugcomment{Accepted 2014 July 7}

\shorttitle{Local instabilities in rotating plasmas}
\shortauthors{Nipoti and Posti}

\begin{document}

\title{On the nature of local instabilities in rotating galactic coronae and cool cores of galaxy clusters}

\author{Carlo Nipoti$^*$ and Lorenzo Posti}
\affil{Department of Physics and Astronomy, Bologna University, viale Berti-Pichat 6/2, 40127 Bologna, Italy}
\email{$^*$ carlo.nipoti@unibo.it}

\begin{abstract}
A long-standing question is whether radiative cooling can lead to
local condensations of cold gas in the hot atmospheres of galaxies and
galaxy clusters.  We address this problem by studying the nature of
local instabilities in rotating, stratified, weakly magnetized,
optically thin plasmas in the presence of radiative cooling and
anisotropic thermal conduction. For both axisymmetric and
non-axisymmetric linear perturbations we provide the general equations
that can be applied locally to specific systems to establish whether
they are unstable and, in case of instability, to determine the kind
of evolution (monotonically growing or over-stable) and the growth
rates of unstable modes. We present results for models of rotating
plasmas representative of Milky Way-like galaxy coronae and cool-cores
of galaxy clusters. It is shown that the unstable modes arise from a
combination of thermal, magnetothermal, magnetorotational and
heat-flux-driven buoyancy instabilities.  Local condensation of cold
clouds tends to be hampered in cluster cool cores, while it is
possible under certain conditions in rotating galactic coronae. If the
magnetic field is sufficiently weak the magnetorotational instability
is dominant even in these pressure supported systems.
\end{abstract}


\section{Introduction}

Extended hot atmospheres are believed to be ubiquitous in massive
virialized systems in the Universe. These virial-temperature gaseous
halos have been detected in X-rays not only in galaxy clusters
\citep{Ros02}, but also in massive elliptical \citep{Mat03} and disc
\citep{Dai12} galaxies. A combination of different observational
findings leads to the conclusion that a corona is present also in the
Milky Way \citep{Mil13}. A long-standing question is whether these
atmospheres are thermally unstable (in the sense of \citealt{Fie65}):
if local thermal instability (TI) occurs, cold gaseous clouds can
condense throughout the plasma; otherwise, substantial cooling can
only happen at the system center, where, however, it is expected to be
opposed by feedback from the central supermassive black hole. This has
important implications for the evolution of galaxy clusters
\citep[e.g.][and references therein]{Mat78,Mcc12} and galaxies
\citep*[e.g.][and references therein]{Mal04,Jou12}.

The evolution of thermal perturbations in astrophysical plasmas,
subject to radiative cooling and thermal conduction, is a complex
physical process \citep{Fie65}, which is influenced by several
factors, such as entropy stratification \citep*{Mal87,Bal89,Bin09},
magnetic fields \citep{Loe90,Bal91,Bal10} and rotation
\citep[][hereafter \citetalias{Nip10}]{Def70,Nip10}.  In the present
paper, which is the follow-up of \citet[][hereafter
  \citetalias{Nip13}]{Nip13}, we focus on the role of rotation in
determining the stability properties of these plasmas, in the presence
of weak magnetic fields. Rotation is clearly expected to be important
in the case of the coronae of disc galaxies \citep{Mar11}, but we
stress that a substantial contribution of rotation could be present
also in the hot gas of galaxy clusters \citep*{Bia13}. Though our
focus is mainly on galaxy and galaxy-cluster atmospheres, it must be
noted that the analysis here presented could be relevant also to the
study of other rotating optically thin astrophysical plasmas, such as
accretion disc coronae \citep{Das13,Li13}.

In \citetalias{Nip13} we have shown that rotating, radiatively cooling
atmospheres with ordered weak magnetic fields (and therefore
anisotropic heat conduction) are unstable to axisymmetric
perturbations, in the sense that there is always at least one growing
axisymmetric mode. The physical implications of this formal result
clearly depend on the nature of this instability, which we try to
address in the present work. In particular, we want to explore whether
the linear instabilities found in \citetalias{Nip13} are
over-stabilities or monotonically growing instabilities, how the
evolution of the instability depends on the properties of the
perturbation and what are the driving physical mechanisms.  We also
extend the linear stability analysis of \citetalias{Nip13} considering
non-axisymmetric linear perturbations.  We analyze the linear
evolution of the instabilities in models of rotating plasmas
representative of Milky Way-like galaxy coronae and cool-cores of
galaxy clusters, comparing the results to those obtained for similar
unmagnetized models, characterized by isotropic heat conduction. The
instabilities found in the present work are interpreted in terms of
well-known instabilities such as the TI, the magnetorotational
instability (MRI, \citealt{BalH91}; see also \citealt{Vel59} and
\citealt{Cha60}), the magnetothermal instability (MTI,
\citealt{Bal00}) and the heat-flux-driven buoyancy instability (HBI,
\citealt{Qua08}).

 The paper is organized as follows. In Section~\ref{sec:goveq} we
 present the relevant magnetohydrodynamic (MHD) equations and we
 define the properties of the unperturbed plasma. The results of the
 linear-perturbation analysis are presented in Section~\ref{sec:axi}
 for axisymmetric perturbations and in Section~\ref{sec:nonaxi} for
 non-axisymmetric perturbations. Section~\ref{sec:con} summarizes and
 concludes.

\section{Governing equations and properties of the unperturbed plasma}
\label{sec:goveq}

A stratified, rotating, magnetized atmosphere in the
presence of thermal conduction and radiative cooling is governed by
the following MHD equations:
\begin{eqnarray}
&&{\de \rho \over \de t}+\grad\cdot(\rho\vv)=0,\label{eq:mass}\\
&&\rho\left[{\partial \vv \over \partial t}+\left(\vv\cdot\grad\right) \vv \right]=-\grad \left(p + {B^2 \over 8\pi}\right) -\rho\grad\Phi + {1 \over 4\pi}(\Bv \cdot \nabla)\Bv,\label{eq:mom}\\
&&{\de \Bv \over \de t} - \rot (\vv \times \Bv) = 0, \label{eq:indu}\\
&&{p\over \gamma-1}\left[{\partial \over \partial  t}+\vv\cdot\grad\right] \ln (p \rho^{-\gamma})=
-\grad\cdot \Qv -\rho\L,\label{eq:ene}
\end{eqnarray}
supplemented by the condition that the magnetic field $\Bv$ is
solenoidal ($\div \Bv = 0$).  Here $\rho$, $p$, $T$ and $\vv$ are,
respectively, the density, pressure, temperature and velocity fields
of the fluid, $\Phi$ is the external gravitational potential (we
neglect self-gravity), $\gamma=5/3$ is the adiabatic index, $\Qv$ is
the conductive heat flux, and $\L=\L(T,\rho)$ is the radiative energy
loss per unit mass of fluid. In a dilute magnetized plasma heat is
significantly transported by electrons only along the magnetic field
lines \citep[see][]{Bra65}, so the conductive heat flux is given by
\begin{eqnarray}
&&
\Qv = -\frac{\chi \Bv\left(\Bv\cdot\nabla\right)T}{ B^2}\qquad 
\mbox{(Anisotropic conduction)},
\label{eq:heatflux}
\end{eqnarray}
where $\chi \equiv \kappa T^{5/2}$ is the Spitzer electron
conductivity with $\kappa\simeq{1.84\times10^{-5}(\ln{\Lambda})}^{-1}
\erg \sminusone \cmminusone \kelvin^{-7/2}$, and $\ln\Lambda$ is the
Coulomb logarithm \citep[][]{Spi62}. In the following we neglect the
weak temperature and density dependence of $\ln\Lambda$, assuming that
$\kappa$ is a constant, so $\chi=\chi(T)\propto T^{5/2}$.  We note
that, even if the medium is magnetized, for simplicity we have assumed
that the pressure is isotropic: in other words, we neglect the
so-called \citet{Bra65} viscosity, i.e. the fact that momentum
transport is anisotropic in the presence of a magnetic field.  Though
this approximation is not rigorously justified (see
\citealt{Kun11,Kun12}; \citealt{Par12}), we adopt it just in the
working hypothesis that anisotropic pressure is not the crucial factor
in determining the thermal stability properties of rotating
plasmas. This is a limitation of the present investigation, which must
be kept in mind in the interpretation of the results.

As we consider rotating fluids, we work in cylindrical coordinates
($R$, $\phi$, $z$), where $R=0$ is the rotation axis.  The unperturbed
plasma is described by time-independent axisymmetric pressure
$\pzero$, density $\rhozero$, temperature $\Tzero$, velocity $\vvzero
= (\vzeroR, \vzerophi, \vzeroz)$ and magnetic field $\Bvzero =
(\BzeroR, \Bzerophi, \Bzeroz)$ satisfying equations
(\ref{eq:mass}-\ref{eq:ene}) with vanishing partial derivatives with
respect to $t$, under the assumption that the unperturbed magnetic
field is subthermal and dynamically unimportant ($\beta\equiv 8\pi
\pzero/\Bzero^2\gg 1$).  Though a stationary solution of the energy
equation~(\ref{eq:ene}) does not necessarily imply
$\rho\L=-\nabla\cdot\Qv$ [for instance, in a classic stationary
  cooling-flow model $\rho\L\neq-\nabla\cdot\Qv$ and
  $\vv\cdot\nabla\ln(p\rho^{-\gamma})\neq 0$], here we restrict for
simplicity to the case in which cooling is balanced by heat conduction
in the unperturbed system. This implies that the unperturbed magnetic
field lines are not isothermal, so $\nabla\Tzero\cdot\bvzero\neq 0$ in
the background plasma (see \citealt{Qua08} for a discussion).  The
unperturbed fluid rotates differentially with angular velocity
$\Omega(R,z) \equiv \vzerophi(R,z)/R$ depending on both $R$ and $z$
(we take $\Omega\geq0$). For simplicity, we assume
$\vzeroR=\vzeroz=0$, which, combined with the assumption that all the
components of the background magnetic fields are time-independent,
implies that the unperturbed system satisfies \citet{Fer37}
isorotation law $\Bvzero\cdot\nabla\Omega=0$ (see \citetalias{Nip13}
for a discussion).  The unperturbed system, though magnetized, obeys
the the Poincar\'e-Wavre theorem \citep[][]{Tas78}, because $\beta\gg
1$, so we can distinguish barotropic [$\pzero=\pzero(\rhozero)$,
  $\Omega=\Omega(R)$] and baroclinic
[$\pzero=\pzero(\rhozero,\Tzero)$, $\Omega=\Omega(R,z)$]
distributions.\\

\section{Axisymmetric perturbations}
\label{sec:axi}

\subsection{Dispersion relation}
\label{sec:disprel}

We linearize the governing equations~(\ref{eq:mass}-\ref{eq:ene}) with
axisymmetric Eulerian perturbations of the form $ F\e^{-\i\omega t +
  \i\kR R + \i\kz z}$, with $|F|\ll|\Fzero|$, where $\Fzero$ is the
unperturbed quantity, $\omega$ is the perturbation frequency, and
$\kR$ and $\kz$ are, respectively, the radial and vertical components
of the perturbation wave-vector.  The linear-perturbation analysis, in
the short wave-length and low frequency (Boussinesq) approximation,
leads to the most general dispersion relation derived in
\citetalias{Nip13}.  {It is convenient to express the dispersion
  relation as a function of $n\equiv-\i\omega$: in terms of $n$, the
  perturbation evolves with a time dependence $F(t)\propto \e^{nt}$,
  where in general $n$ is a complex number.}  Therefore, stable modes
are those with ${\rm Re}(n)\leq 0$ and unstable modes those with ${\rm
  Re}(n)> 0$. Among the unstable modes we have monotonically growing
unstable modes, with ${\rm Im}(n)=0$, and over-stable modes, with
${\rm Im}(n)\neq0$, in which the perturbation oscillates with growing
amplitude.  The dispersion relation of \citetalias{Nip13} (equation 32
in that paper) can be written in dimensionless form as
\begin{eqnarray}
&&
\ntil^5
+ \omegadtil \ntil^4
+ \left[\omegaBVsqtil + \omegarotsqtil + 2 \omegaAsqtil\right]\ntil^3
+ \left[(\omegarotsqtil + 2 \omegaAsqtil)\omegadtil + \omegaAsqtil\omegacmagtil \right]\ntil^2
+ \omegaAsqtil \left( \omegaAsqtil 
+ \omegaBVsqtil + \omegarotsqtil - 4\frac{\kz^2}{k^2}+\omegacphisqtil\right)\ntil
\nonumber\\
&&
\qquad
+ \omegaAsqtil\left[ \left(\omegaAsqtil + \omegarotsqtil - 4\frac{\kz^2}{k^2} \right)\omegadtil  + \omegaAsqtil\omegacmagtil\right] = 0\qquad \mbox{(Magnetized)}.
\label{eq:disp_np13}
\end{eqnarray}
In the above equation $\ntil\equiv n/\Omega$,
$\omegadtil\equiv\omegacatil + \omegathtil$ (where
$\omegacatil\equiv\omegaca/\Omega$ and
$\omegathtil\equiv\omegath/\Omega$),
$\omegaBVsqtil\equiv\omegaBVsq/\Omega^2$,
$\omegarotsqtil\equiv\omegarotsq/\Omega^2$, $ \omegaAsqtil\equiv
\omegaAsq/\Omega^2$, $\omegacmagtil\equiv\omegacmag/\Omega$ and
$\omegacphisqtil \equiv \omegacphisq/\Omega^2$, where
\begin{eqnarray}
&&\omegaca\equiv \frac{(\kv \cdot \bvzero)^2}{k^2}\omegac
\end{eqnarray}
is the anisotropic thermal conduction frequency, 
\begin{eqnarray}
&&\omegac\equiv\left({\gamma-1\over\gamma}\right){k^2 \chi(\Tzero) \Tzero\over \pzero}
\end{eqnarray}
is the isotropic thermal conduction frequency,
\begin{eqnarray}
&&\omegath \equiv-\left({\gamma-1\over\gamma}\right)\frac{\rhozero}{\pzero}
\left[\L(\rhozero,\Tzero)+\rhozero \Lrho(\rhozero,\Tzero) 
-\Tzero \LT(\rhozero,\Tzero) \right]
\end{eqnarray}
is the thermal-instability frequency\footnote{This is a general
  definition of $\omegath$, which applies also in the case in which
  $\rho\L\neq -\nabla\cdot\Qv$ in the unperturbed system: it can be
  shown that the first term in square brackets is negligible if
  cooling is balanced by heat conduction in the background fluid.},
\begin{eqnarray}
&&\omegaBVsq \equiv-\kzksq{\Dm \pzero \over \rhozero\gamma}\Dm  \ln \pzero \rhozero^{-\gamma}
\end{eqnarray}
is the Brunt-V\"ais\"al\"a or buoyancy frequency squared [we have
  introduced the differential operator $\Dm\equiv
{({\kR}/{\kz})}{\partial}/{\partial z}-{\partial}/{\partial R}$],
\begin{eqnarray}
&&\omegarotsq\equiv  -\kzksq{1\over R^3} \Dm({R^4 \Omega^2})
\end{eqnarray}
is the angular-momentum gradient frequency squared,
\begin{eqnarray}
&&\omegaAsq\equiv (\kv\cdot \vvA)^2=\frac{(\kv\cdot{\Bvzero})^2}{4\pi\rhozero}
\end{eqnarray}
is the Alfv\'en frequency squared, and
\begin{eqnarray}
&&\omegacmag \equiv 
-\omegac
\frac{4\pi\pzero}{\Bzero^2}
\kzksq
\frac{\Dm \ln \pzero}{k^2}
\left[
\Dm \ln \Tzero
-2\left(\nabla \ln \Tzero\cdot \bvzero\right)
\left(\frac{\kR}{\kz}{\bzeroz}-\bzeroR\right)
\right]
\end{eqnarray}
and
\begin{eqnarray}
&& \omegacphisq \equiv 
\omegac\Omega
\frac{8\pi \pzero}{\Bzero^2}
\kzksq
\frac{\Dm\ln \pzero}{k^2} 
(\nabla \ln \Tzero\cdot \bvzero)\bzerophi
\end{eqnarray}
are other two frequencies associated to thermal conduction
mediated by the magnetic field.

We are interested in comparing the thermal stability properties of
plasmas with ordered magnetic field to those of unmagnetized fluids
with similar properties, in which thermal conduction is isotropic.  In
this case the governing equations are the usual hydrodynamic
equations, which can be obtained from equations~(\ref{eq:mass}),
(\ref{eq:mom}) and (\ref{eq:ene}), imposing $\Bv=0$ and conductive
heat flux (\citealt{Spi62})
\begin{eqnarray}&&
\Qv = - f \kappa T^{5/2} \nabla T,\qquad \mbox{(Isotropic conduction)},
\label{eq:tangledtc}
\end{eqnarray}
where we allow for the possibility that thermal conduction is
suppressed to a fraction $f$ of the classical Spitzer's value
\citep[][]{Bin81}, so that the unmagnetized case can also represent a
simple model of system with tangled magnetic fields. The dispersion
relation is the one derived in \citetalias{Nip10} (equation 16 in that
paper), which we write here as a function of $\ntil=n/\Omega$:
\begin{eqnarray}
&&
\ntil^3 +\ntil^2\omegadtil
+(\omegaBVsqtil+\omegarotsqtil)\ntil+\omegarotsqtil\omegadtil=0\qquad \mbox{(Unmagnetized)},
\label{eq:disp_n10}
\end{eqnarray}
where now $\omegadtil\equiv f\omegactil + \omegathtil$, with
$\omegactil\equiv\omegac/\Omega$.

\begin{table}
\begin{center}
\newcommand\T{\rule{0pt}{4ex}}
\caption{\label{tab:dimlesspar}}
{\small
\begin{tabular}{lllllllllllllll}
\tableline\tableline
Name &  $\GammaOmegaR$ & $\GammaOmegaz$ & $\GammapR$ & $\Gammapz$ & $\GammaTR$ & $\czerotil$ & $\omegathtil$ & $\omegaczerotil$ & $\beta$ & $\bzerophi$ & $\bzeroR$ & $\bzeroz$  & $\GammaTz$ & $\GammaTp$ \\
\tableline
   MWG-bt &   -0.944 &   0 &   -0.321 &   -0.565 &   -0.042 &    1.238 &   -0.052 &    0.010 &   29.15 &    0 &   0 &    1.000 &   -0.074 &    0.130 \\
 MWG-bt-az &   -0.944 &   0 &   -0.321 &   -0.565 &   -0.042 &    1.238 &   -0.052 &    0.010 &   29.15 &    0.950 &   0 &    0.312 &   -0.074 &    0.130 \\
    MWG-bc &   -0.944 &   -0.142 &   -0.321 &   -0.565 &   -0.042 &    1.238 &   -0.052 &    0.010 &   29.15 &    0 &   -0.148 &    0.989 &    0.503 &   \\
 MWG-bc-az &   -0.944 &   -0.142 &   -0.321 &   -0.565 &   -0.042 &    1.238 &   -0.052 &    0.010 &   29.15 &    0.950 &   -0.046 &    0.309 &    0.503 &   \\
    CCC-bt &   -0.868 &   0 &   -0.249 &   -0.324 &    0.173 &    1.834 &   -0.061 &    0.128 &  77.73 &    0 &   0 &    1.000 &    0.225 &   -0.695 \\
    CCC-bc &   -0.868 &   -0.130 &   -0.249 &   -0.324 &    0.173 &    1.834 &   -0.061 &    0.128 &  77.73 &    0 &   -0.148 &    0.989 &    0.537 &   \\
\tableline
\end{tabular}
}
\tablecomments{List of dimensionless parameters of the models (see
    Section~\ref{sec:param}). The first ten parameters are
    independent: logarithmic slopes of the radial ($\GammaOmegaR$) and
    vertical ($\GammaOmegaz$) angular velocity gradients, of the
    radial ($\GammapR$) and vertical ($\Gammapz$) pressure gradients,
    and of the radial temperature gradient ($\GammaTR$), normalized
    sound speed ($\czerotil$), normalized thermal-instability
    ($\omegathtil$) and thermal-conduction ($\omegaczerotil$)
    frequencies, thermal to magnetic pressure ratio ($\beta$), and
    normalized azimuthal magnetic field component ($\bzerophi$). The
    last four parameters depend on the first ten: normalized radial
    ($\bzeroR$) and vertical ($\bzeroz$) magnetic field components,
    logarithmic slope of the vertical temperature gradient
    ($\GammaTz$) and logarithmic slope of the temperature variation
    with respect to pressure ($\GammaTp$; defined only for barotropic
    distributions).\\}
\end{center}
\end{table}

\subsection{Parameters}
\label{sec:param}

We show here that the coefficients of the dispersion
relation~(\ref{eq:disp_np13}) are fully defined if the values of the
axisymmetric perturbation wave-vector $\kv=(\kR,0,\kz)$ and of ten
dimensionless parameters (depending on the physical properties of the
unperturbed plasma) are specified. {For this purpose, it is
  convenient to rewrite the characteristic frequencies appearing in
  the coefficients as follows:
\begin{eqnarray}
&&
\omegacatil =\frac{k^2R^2}{1+x^2} (\bzeroz+x\bzeroR)^2\omegaczerotil,
\label{eq:omegacatil}\\
&&
\omegactil=k^2R^2\omegaczerotil,
\label{eq:omegactil}\\
&&
\omegacmagtil=
-\frac{\beta}{2(1+x^2)}
\omegaczerotil
(x\Gammapz-\GammapR)
\Big[
(x\GammaTz-\GammaTR)
-2\left(\GammaTR\bzeroR+\GammaTz\bzeroz\right)
\left(x\bzeroz-\bzeroR\right)
\Big],
\label{eq:omegacmagtil}\\
&&
\omegacphisqtil=
\frac{\beta}{1+x^2}
\omegaczerotil
(x\Gammapz-\GammapR)
\left(\GammaTR\bzeroR+\GammaTz\bzeroz\right)
\bzerophi,
\label{eq:omegacphi}\\
&&
\omegaBVsqtil =-\frac{\czerotil^2}{1+x^2}(x\Gammapz-\GammapR)\left[x\GammaTz-\GammaTR -\frac{\gamma-1}{\gamma}\left(x\Gammapz-\GammapR\right)\right],
\label{eq:omegaBVtil}\\
&&
\omegarotsqtil 
=\frac{2}{1+x^2}\left(\GammaOmegaR+2-x\GammaOmegaz\right),
\label{eq:omegarottil}\\
&&
\omegaAsqtil=\frac{2\czerotil^2}{\beta}\frac{k^2R^2}{1+x^2}(x\bzeroR+\bzeroz)^2,
\label{eq:omegaAtil}\\
&&
\omegathtil \equiv-\frac{R}{\vzerophi}\left({\gamma-1\over\gamma}\right)\frac{\rhozero}{\pzero}
\left[\L(\rhozero,\Tzero)+\rhozero \Lrho(\rhozero,\Tzero) 
-\Tzero \LT(\rhozero,\Tzero) \right],
 \label{eq:omegathtil}
\end{eqnarray}
where we have introduced the dimensionless parameters $x\equiv \kR/\kz
$, $\bvzero\equiv \Bvzero/\Bzero$, $\czerotil\equiv\czero/\vzerophi$
($\czero$ is the isothermal sound speed), 
\begin{eqnarray}
&&\GammapR\equiv\frac{\de \ln \pzero }{ \de \ln R},\qquad
\Gammapz\equiv\frac{R}{z}\frac{\de \ln \pzero}{\de \ln |z|},\\
&&\GammaTR\equiv\frac{\de \ln \Tzero}{\de \ln R}, \qquad
\GammaTz\equiv\frac{R}{z}\frac{\de \ln \Tzero}{ \de \ln |z|},\\
&&\GammaOmegaR\equiv\frac{\de \ln \Omega}{ \de \ln R},\qquad
\GammaOmegaz\equiv\frac{R}{z}\frac{\de \ln \Omega}{\de \ln |z|}
\end{eqnarray}
and
\begin{eqnarray}
&&\omegaczerotil\equiv \frac{\gamma-1}{\gamma}\frac{\chi(\Tzero) \Tzero}{ \vzerophi R \pzero}.
\end{eqnarray}}
In summary, for given $\kv$, the dispersion
relation~(\ref{eq:disp_np13}) for a magnetized plasma is completely
defined by the choice of the following ten free parameters, depending
only on the properties of the unperturbed plasma: $\GammaOmegaR$ and
$\GammaOmegaz$ (logarithmic slopes of the radial and vertical angular
velocity gradients), $\GammapR$ and $\Gammapz$ (logarithmic slopes of
the radial and vertical pressure gradients), $\GammaTR$ (logarithmic
slope of the radial temperature gradient), $\czerotil$ (sound to
rotation speed ratio), $\omegathtil$ and $\omegaczerotil$ (normalized
thermal-instability and thermal-conduction frequencies), $\beta$
(thermal to magnetic pressure ratio) and $\bzerophi$ (normalized
azimuthal magnetic field component).  {The other three parameters
appearing in the definitions of the coefficients of the dispersion
relations ($\bzeroR$, $\bzerophi$ and $\GammaTz$) are not independent
of the first ten listed above: $\bzeroR$ and $\bzeroz$ are determined
(modulo the sign) when $\GammaOmegaz$, $\GammaOmegaR$ and $\bzerophi$
are given, because $\GammaOmegaz\bzeroz=-\GammaOmegaR\bzeroR$
(isorotation condition) and $\bzeroR^2+\bzeroz^2+\bzerophi^2=1$;
$\GammaTz$ is determined by the vorticity equation (derived from the
momentum equations), which can be written in dimensionless form as
\begin{eqnarray}
&&\GammaOmegaz=\frac{\czerotil^2}{2}\left(\GammaTz\GammapR-\GammaTR\Gammapz\right).
\label{eq:vort}
\end{eqnarray}
In the special case of barotropic distributions
$\GammaTR=\GammaTp\GammapR$ and $\GammaTz=\GammaTp\Gammapz$, where
\begin{eqnarray}
&&\GammaTp\equiv\frac{\d \ln \Tzero}{ \d \ln \pzero},
\end{eqnarray}
so $\GammaOmegaz=0$ (see equation~\ref{eq:vort}) and $\bzeroR=0$ (and
therefore $\bzeroz^2+\bzerophi^2=1$), because of the isorotation
condition $\bzeroR\GammaOmegaR=0$.} We note that, at fixed $\GammaTR$,
$\GammaTz$ and $\bvzero$, the requirement that in the unperturbed
system cooling is balanced by heat conduction
$\left(\rho\L=-\nabla\cdot\Qv\right)$ can be fulfilled, for given
$\rho\L$, by fixing the values of the magnetic field gradient and of
the second spatial derivatives of the temperature. The dispersion
relation~(\ref{eq:disp_n10}) for an unmagnetized fluid is fully
characterized by the parameters $\GammaOmegaR$, $\GammaOmegaz$,
$\GammapR$, $\Gammapz$, $\GammaTR$, $\czerotil$, $\omegathtil$,
$\omegaczerotil$ and $f\leq1$.

\subsection{Model astrophysical plasmas}
\label{sec:mod}

\begin{table}
\caption{
\label{tab:physpar}}
\begin{center}
\begin{tabular}{lllllll}
\tableline\tableline
Name  & $\Tzero/K$  & $\nezero/\cmminusthree$ & $R/\kpc$ &  $\vzerophi/\kms$ & $Z/\Zsun$ & $B_0/\mu G$ \\ 
\tableline
 MWG & $2.15\times10^6$ & $8.58\times10^{-4}$ &   16.92 &  140.2 &    0.03  &    0.5 \\
 CCC & $7.96\times10^7$ & $2.47\times10^{-2}$ &   92.00 &  576.0 &    0.30  &   10 \\
\tableline
\end{tabular}
\tablecomments{Examples of combinations of physical parameters (see
    Section~\ref{sec:mod}) giving $\czerotil$, $\omegathtil$,
    $\omegaczerotil$ and $\beta$ as in Table~\ref{tab:dimlesspar}.
    $\Tzero$ is the gas temperature, $\nezero$ is the electron number
    density, $R$ is the local radial coordinate (representing the
    characteristic physical size of the system), $Z$ is the metallicity
    and $\Bzero$ is the magnetic field modulus.  Here we estimate
  $\L(\rhozero,\Tzero)$, appearing in $\omegathtil$, using the
  collisional ionization equilibrium cooling function of
  \citet{Sut93}.\\}
\end{center}
\end{table}

To study the linear evolution of given modes, we must specify the
numerical values of the coefficients of the dispersion relations
(\ref{eq:disp_np13}) and (\ref{eq:disp_n10}). Given the number of
parameters (see Section~\ref{sec:param}), it is clear that an
exploration of parameter space is prohibitive. However, it is useful
to present results for a few illustrative cases. We focus on two
reference sets of models, which are meant to be representative of the
physical conditions in the cool core of a massive galaxy cluster (set
of models CCC) and in the corona of a Milky Way-like galaxy (set of
models MWG).  In all cases the hot gas is assumed to be rotating
differentially, and we consider both barotropic [$\Omega=\Omega(R)$,
  so $\GammaOmegaz=0$; models CCC-bt and MWG-bt] and baroclinic
[$\Omega=\Omega(R,z)$, so $\GammaOmegaz\neq0$; models CCC-bc and
  MWG-bt] distributions. In these models the magnetic field has no
azimuthal component ($\bzerophi=0$), but we also present results for
two models with $\bzerophi\neq0$ (models MWG-bt-az and MWG-bc-az).
Physically, the main differences between the two sets are that
temperature is higher (i.e. conduction more efficient) and rotation
less important in models CCC than in models MWG, and that the radial
temperature gradient is positive in models CCC and negative in models
MWG. The values of the dimensionless parameters for all the models are
reported in Table~\ref{tab:dimlesspar}.  For each of the above models
we construct a corresponding unmagnetized model that has the same
values of the parameters $\GammaOmegaR$, $\GammaOmegaz$, $\GammapR$,
$\Gammapz$, $\GammaTR$, $\czerotil$, $\omegathtil$ and
$\omegaczerotil$.

Some of the parameters listed in Table~\ref{tab:dimlesspar} are purely
``geometrical'': those indicated with $\Gamma$ describe the local
density, pressure, temperature and velocity gradients, while $\bv$
describes the geometry of the local magnetic field. The others
($\czerotil$, $\omegathtil$, $\omegaczerotil$, $\beta$) are
combinations of physical parameters of the unperturbed system, but
clearly they are not associated to specific values of the parameters
in physical units. For instance, a given numerical value of
$\czerotil=\czero/\vzerophi$ can be obtained with an arbitrary value
of the sound speed $\czero$ (in physical units), provided that the
rotation speed $\vzerophi$ is opportunely rescaled. In this sense, our
models are not univocally associated to given physical scales,
temperatures, densities, velocities and magnetic fields. Nevertheless,
for illustrative purposes, it is useful to give examples of
combinations of physical parameters (relevant to astrophysical
applications) that produce the values of the dimensionless parameters
reported in Table~\ref{tab:dimlesspar}. The quantities entering the
definition of $\czerotil$, $\omegathtil$, $\omegaczerotil$ and $\beta$
are the local temperature $\Tzero$, the local density $\rhozero$ (or,
alternatively, the local electron number density $\nezero$), the local
radial coordinate $R$ (representing the characteristic physical size
of the system), the local rotation speed $\vzerophi$, the metallicity
$Z$ (because of the dependence of $\L$, an therefore of $\omegathtil$,
on $Z$) and the modulus of the local magnetic field $\Bzero$. Two
examples of combinations of the values of these parameters are
reported in Table~\ref{tab:physpar}: it is clear that model CCC can be
interpreted as the cool core of a massive cluster of galaxies and
model MWG as a galactic corona. In fact, the values of the
dimensionless parameters of our sets of models MWG and CCC were
inspired, respectively, by the Milky Way-like corona models of
\citet{Bin09} and by the cool-core cluster models of \citet{Bia13}.
In global models of hot atmospheres (such as those of \citealt{Bin09}
and \citealt{Bia13}) the physical parameters (and then the detailed
stability properties) are clearly position-dependent. In the present
work we do not explore global models, but we just take specific values
of the parameters that are representative of the physical conditions
in typical galaxies or clusters.  However, we stress that once a
global model of a rotating atmosphere is given (either analytic or
obtained with a numerical simulation), its thermal stability can be
studied at any point of the system by solving the dispersion
relations~(\ref{eq:disp_np13}) and (\ref{eq:disp_n10}) with the values
of the coefficients given by the physical properties at that point.

\subsection{Results for axisymmetric perturbations}
\label{sec:resaxi}

Given a plasma model, we solved numerically the dispersion
relations~(\ref{eq:disp_np13}) and (\ref{eq:disp_n10}) for a large
number of axisymmetric wave-vectors, using the \idl (Interactive Data
Language) routine \fzroots.  The analysis of the roots as functions of
the wave-vector components allows us to identify domains of stability,
over-stability and monotonic instability.  In order to understand the
nature of the instabilities it is important to assess the role of
different physical mechanisms, such as radiative cooling, rotation and
anisotropic heat conduction. This task is by no means easy, because
the different modes are entangled and the high order of the dispersion
relation makes it difficult to isolate the various
contributions. Nevertheless, inspecting the behavior and the growth
rates of the individual branches of the solutions as functions of the
wave-number and comparing the results with those obtained for simpler
configurations (for instance in the absence of radiative cooling,
rotation or magnetic field), we were able to identify the different
branches in terms of combinations of well known modes (TI, MRI, MTI,
HBI, rotation, buoyancy and Alfv\'en modes).

\subsubsection{Domains of stability, over-stability and monotonically growing instability}
\label{sec:dom}

\begin{figure}
\centerline{\psfig{file=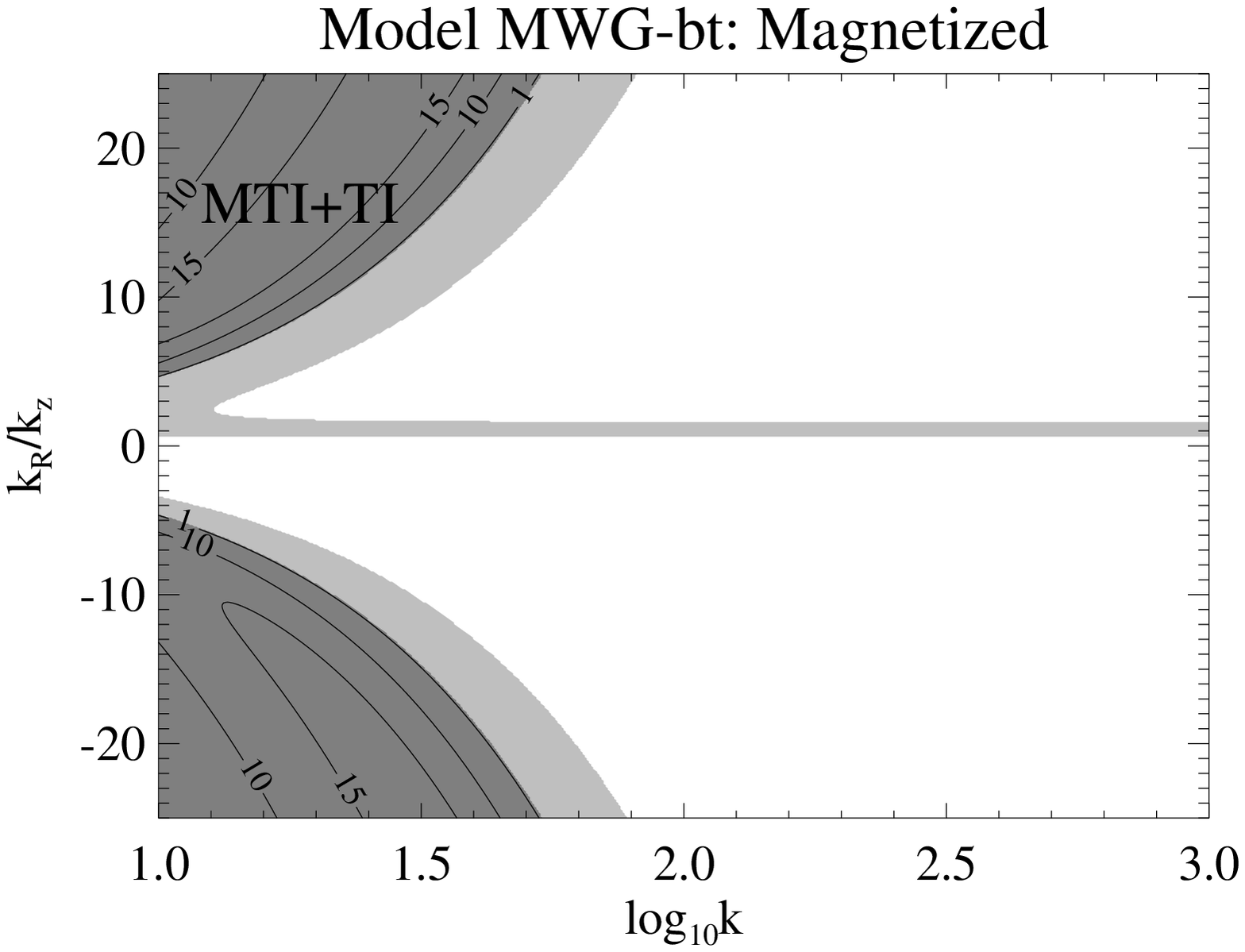,width=0.5\hsize}\psfig{file=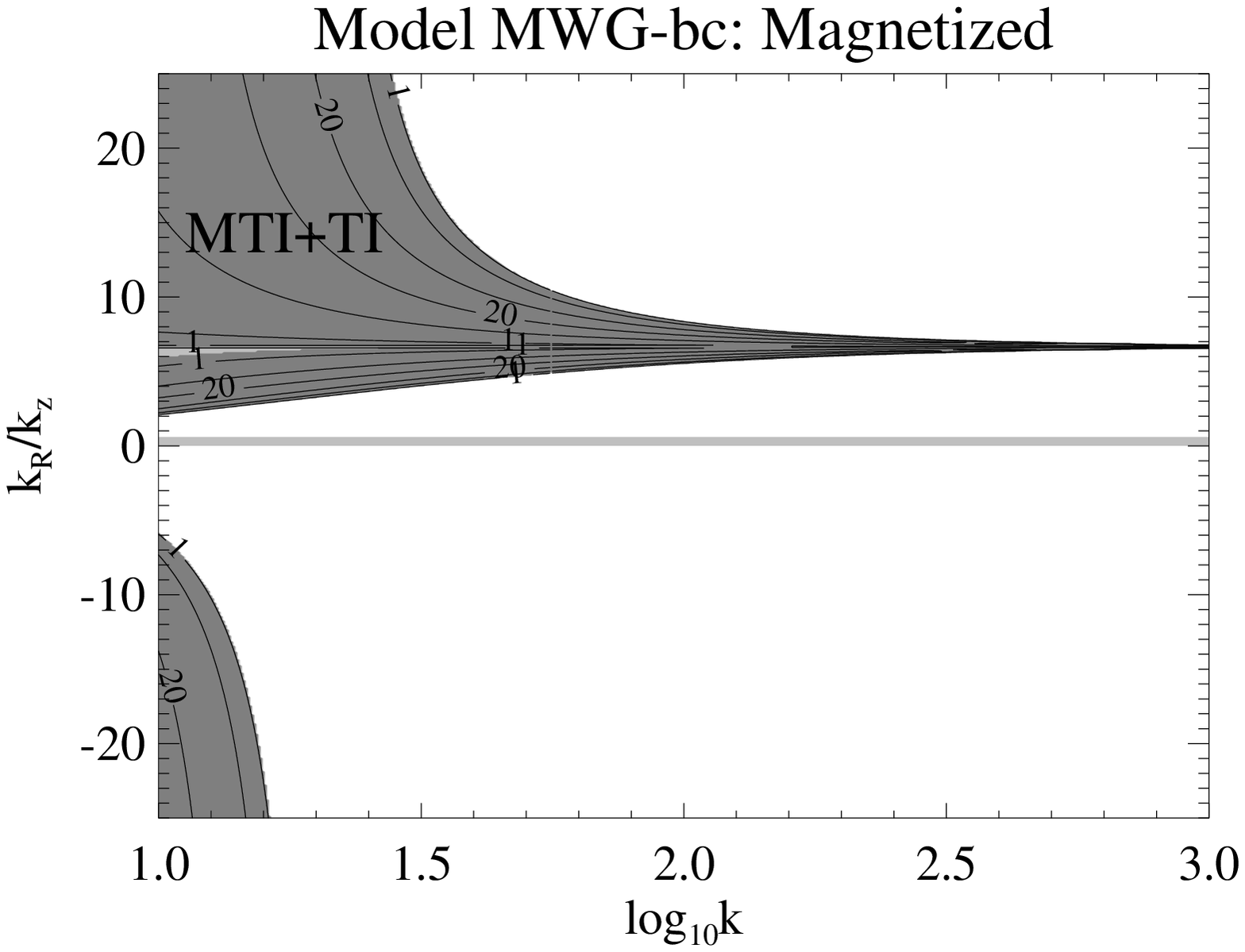,width=0.5\hsize}}
\caption{Domains of stability (white), monotonic instability (dark
  grey) and over-stability (light grey) in the plane $\kR/\kz$ (ratio
  of radial and vertical wave-vector components) vs $k$ (wave-number)
  for barotropic (left-hand panel) and baroclinic (right-hand panel)
  Milky-Way like galaxy models with with $\bzerophi=0$ (see Table
  \ref{tab:dimlesspar}). Representative contours of constant growth
  rate (labeled in units of $10^{-3}\Omega^{-1}$) are shown for the
  monotonically unstable modes. In both panels the dominant
  instabilities are due to a combination of magnetothermal and thermal
  modes (label MTI+TI). Here $k$ is in units of $R^{-1}$.\\}
\label{fig:dom_MWG}
\end{figure}

\begin{figure}
\centerline{\psfig{file=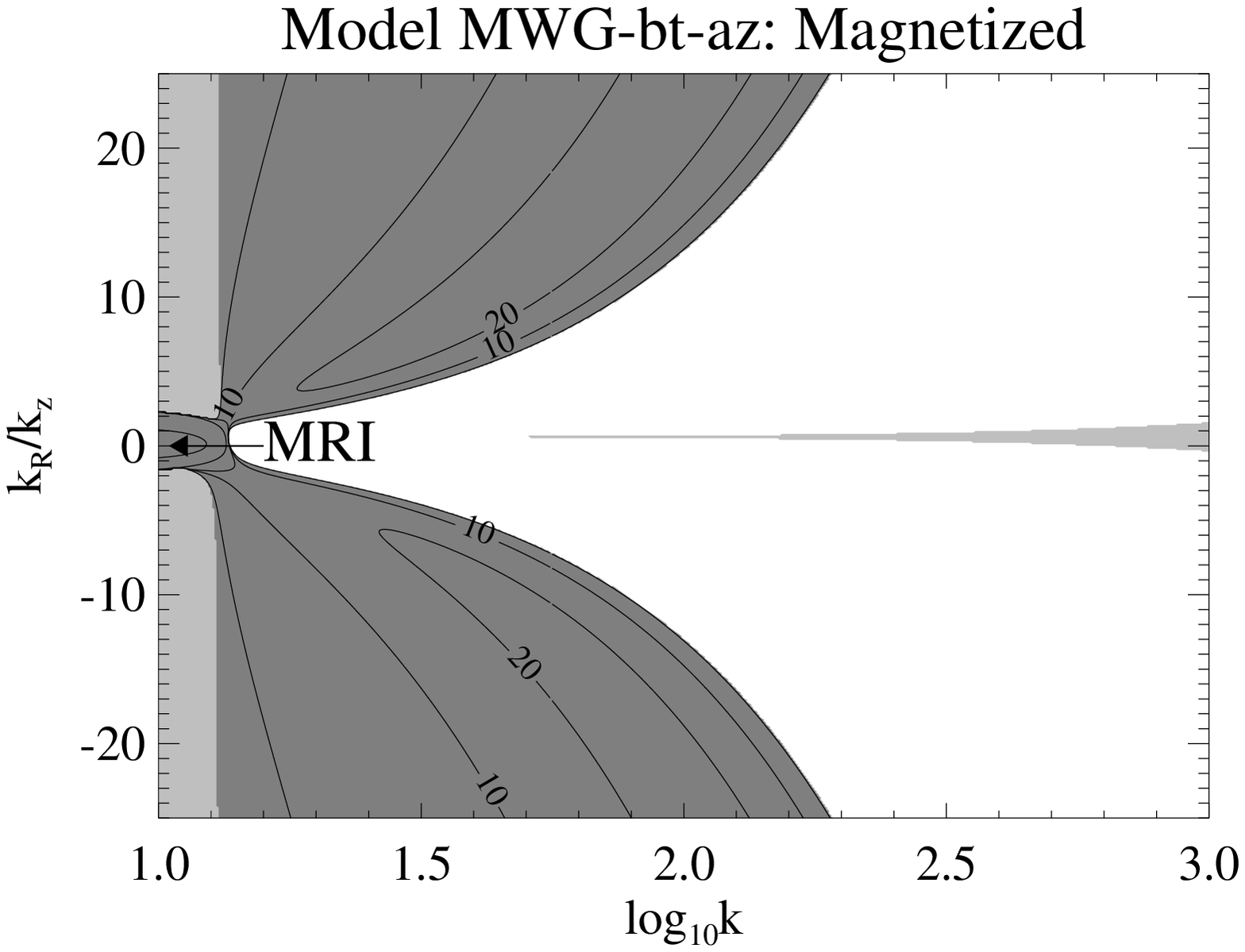,width=0.5\hsize}\psfig{file=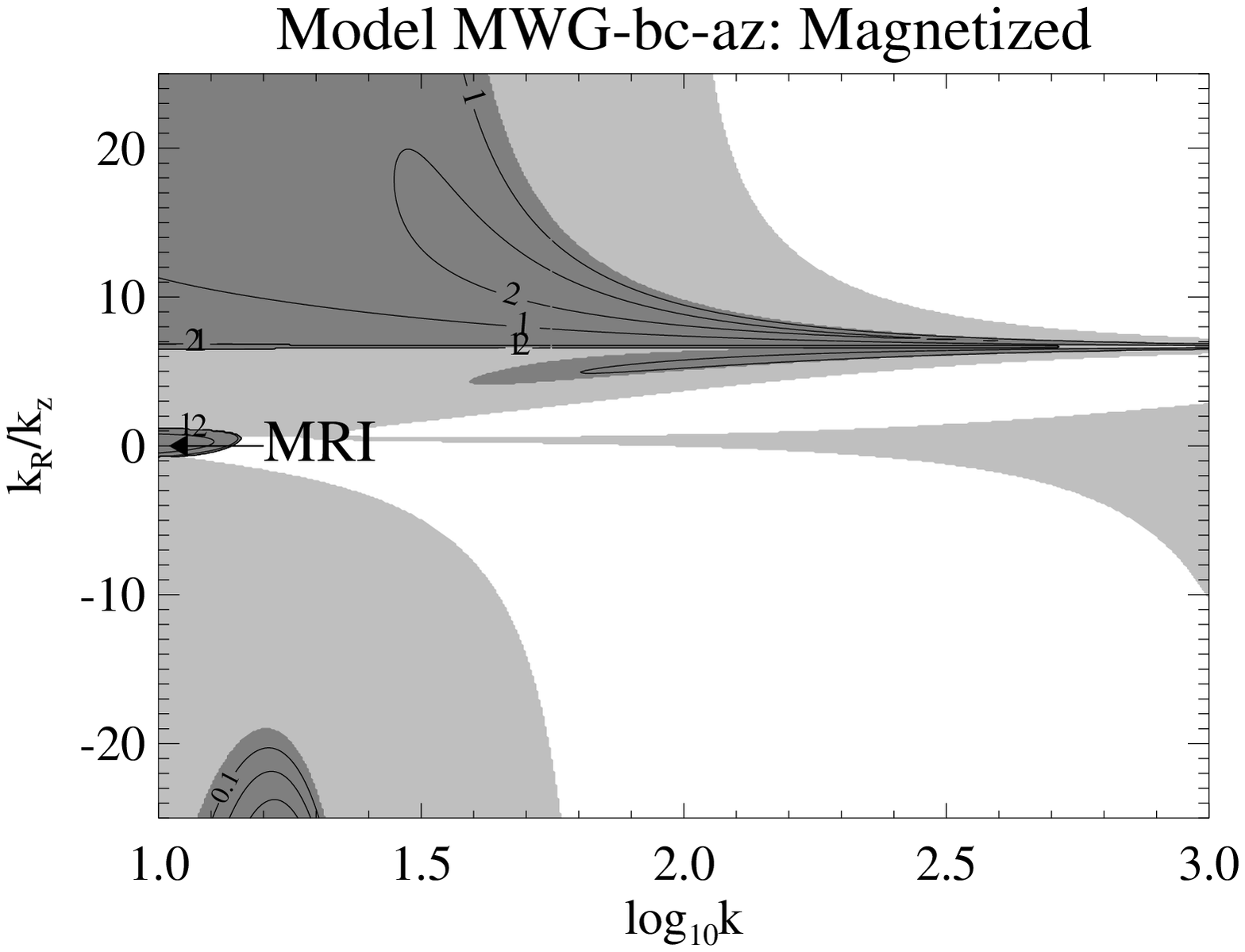,width=0.5\hsize}}
\caption{Same as Fig.~\ref{fig:dom_MWG}, but for barotropic (left-hand
  panel) and baroclinic (right-hand panel) Milky-Way like galaxy
  models with $\bzerophi\neq0$ (see Table \ref{tab:dimlesspar}). In
  both panels the fastest growing monotonic instabilities (growth rate
  up to $\approx 440$ in units of $10^{-3}\Omega^{-1}$) are
  magnetorotational modes with $|\kR/\kz|\lesssim 1$ and $\log_{10}
  k\lesssim 1.1$ (label MRI).\\}
\label{fig:dom_MWG_az}
\end{figure}

\begin{figure}
\centerline{\psfig{file=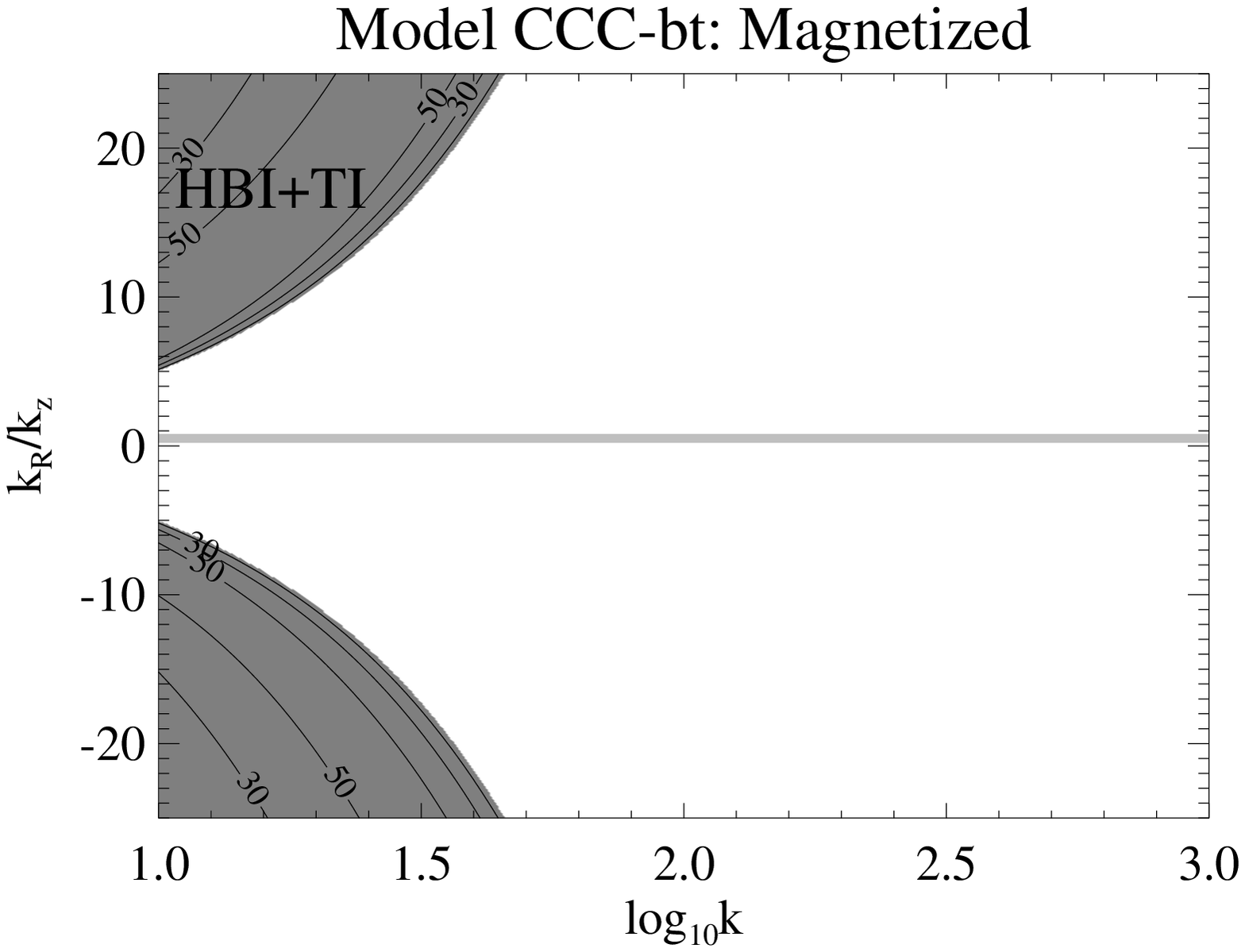,width=0.5\hsize}\psfig{file=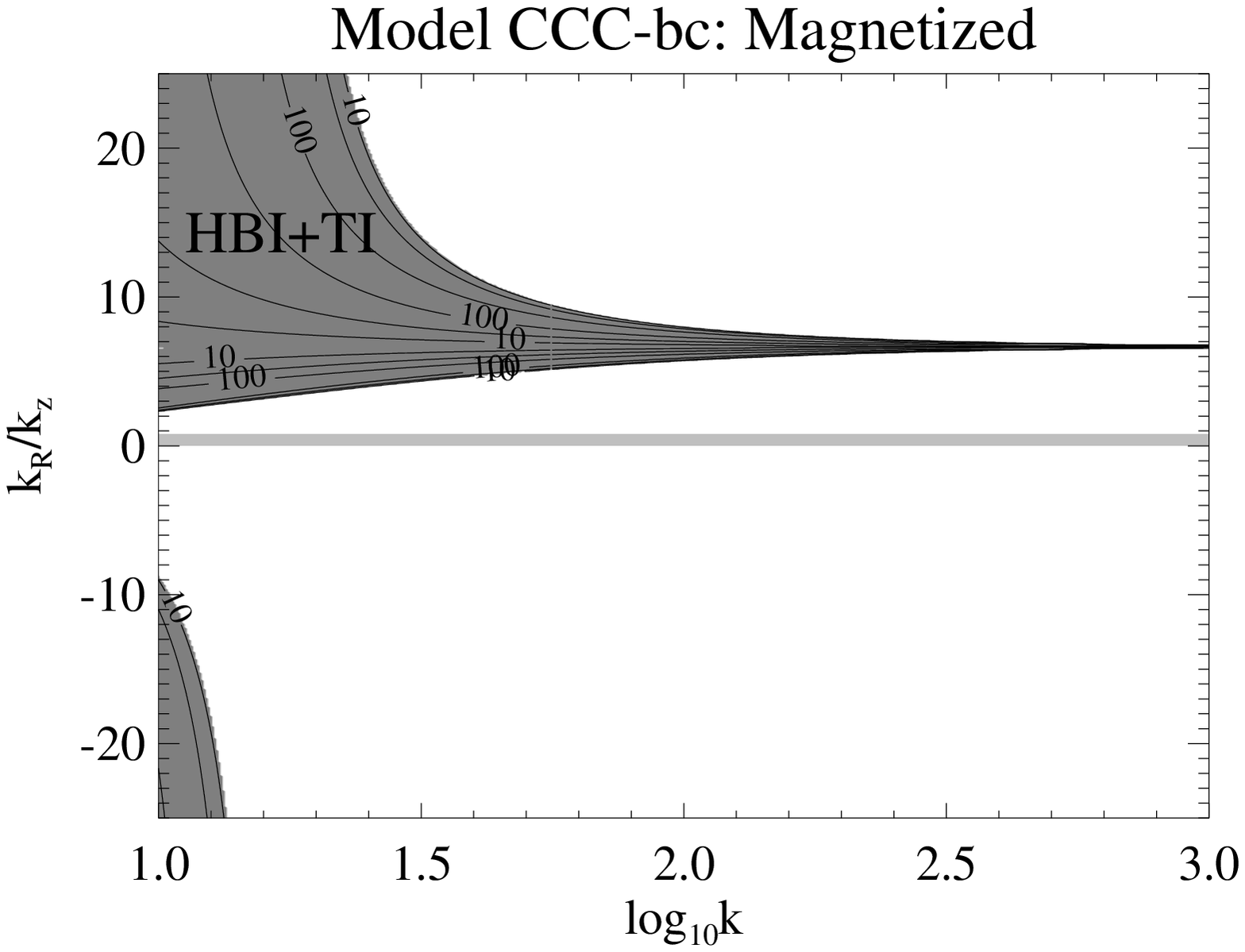,width=0.5\hsize}}
\caption{Same as Fig.~\ref{fig:dom_MWG}, but for barotropic (left-hand
  panel) and baroclinic (right-hand panel) cool-core cluster models
  (see Table \ref{tab:dimlesspar}). In both panels the dominant
  instabilities are due to a combination of heat-flux driven buoyancy
  and thermal modes (label HBI+TI).\\}
\label{fig:dom_CCC}
\end{figure}

Here we discuss the domains, in wave-vector space, of stability,
over-stability and monotonic instability against axisymmetric
perturbations for the plasma models described in
Section~\ref{sec:mod}.  The results for the Milky-Way like galaxy
models MWG-bt (barotropic) and MWG-bc (baroclinic) with $\bzerophi=0$
are shown in Fig.~\ref{fig:dom_MWG}: in these plots the stability
domains are represented in the space $k$-$x$, where $k$ is the
wave-number of the perturbation and $x\equiv\kR/\kz$ is the ratio of
the radial and vertical components of the wave-vector.  In particular,
we sampled the intervals $10\leq k R\leq1000$ and $-25\leq x\leq
25$. From Fig.~\ref{fig:dom_MWG} it is apparent that in all cases
there are regions in the space $k$-$x$ corresponding to monotonically
unstable and over-stable modes, typically for small $k$ (i.e., long
wave-length) and large $|x|$ (i.e. almost vertical wave-crests). In
the plot representative contours of constant growth rate are shown for
the monotonically unstable modes, which in this case are due to a
combination of the MTI and the TI (see Section~\ref{sec:nat}).  In the
baroclinic MWG model (right-hand panel in Fig.~\ref{fig:dom_MWG})
monotonic instability occurs also at short wave-lengths (a region of
stability or over-stability for barotropic models) for $x\approx
-\bzeroz/\bzeroR\simeq 6.7$: physically this is due to the fact that
for those modes the projection of the magnetic field onto the wave
vector is close to zero, so the fluid basically behaves as if it were
unmagnetized ($\omegaA\approx0$) and with no heat conduction
($\omegaca\approx 0$; see also \citetalias{Nip13}). The driving term
of these growing disturbances is $\omegath$, so the instability is
essentially a TI (see Section~\ref{sec:nat}).  The domains of
stability and instability in models with $\bzerophi\neq0$
(Fig.~\ref{fig:dom_MWG_az}) present a region of instability for low
$k$ and $|x|\lesssim 1$, which is not present in the corresponding
models with $\bzerophi=0$ (Fig.~\ref{fig:dom_MWG}): we will show in
Section~\ref{sec:nat} that these unstable modes are dominated by the
MRI.

The stability and instability domains of the barotropic (CCC-bt) and
baroclinic (CCC-bc) cluster cool core models are qualitatively similar
to those of corresponding MWG models, as it is apparent by comparing
Fig.~\ref{fig:dom_CCC} with Fig.~\ref{fig:dom_MWG}.  Quantitatively,
in the $k$-$x$ space, the regions of monotonic instability are less
extended in the CCC models than in the MWG models, mainly because
damping by thermal conduction is more effective in the
higher-temperature plasma of cool cores than in galactic coronae.
Moreover, as we will discuss in Section~\ref{sec:nat}, the driving
instability is different in the two cases (HBI in models CCC, MTI in
models MWG).

It is interesting to compare the above results for magnetized plasma
models to those obtained for the corresponding unmagnetized models
(see Section~\ref{sec:mod}), for which the stability domains can be
expressed analytically in terms of the quantities $\omegadtil$,
$\omegarotsqtil$ and $\omegaBVsqtil$ (see \citetalias{Nip10}, figure 1
in that paper).  The behavior of the barotropic models with no
magnetic field is very simple: all modes with $k>\kmin$
(i.e. sufficiently short wave-length perturbations) are either stable
or over-stable (due to effective heat-conduction damping), while for
$k<\kmin$ we have monotonic instability: for instance, for model
MWG-bt $\kmin\simeq 22 (f/0.01)^{-1/2}$, where $f\leq1$ is the
thermal-conduction suppression factor\footnote{In the unmagnetized
  case, the condition for heat-conduction damping is
  $\omegadtil=f\omegactil + \omegathtil\geq0$, where, for fixed
  unperturbed physical parameters, $\omegactil\propto k^2$ and
  $\omegathtil=const$. The minimum stable wave-number $\kmin$
  corresponds to $\omegadtil=0$, so $\kmin^2f=const$.}.  Baroclinic
models are characterized also by a critical wave-vector component
ratio $\xcrit$ (for instance, $\xcrit\simeq-7$ for model MWG-bc): for
$x>\xcrit$ the behavior is similar to that of the corresponding
barotropic models, while for $x<\xcrit$ the system is rotationally
unstable ($\omegarotsq<0$), which, combined with the Field criterion,
gives over-stability for $k<\kmin$ and instability for $k>\kmin$ (see
\citetalias{Nip10}).

\begin{figure}
\centerline{
\psfig{file=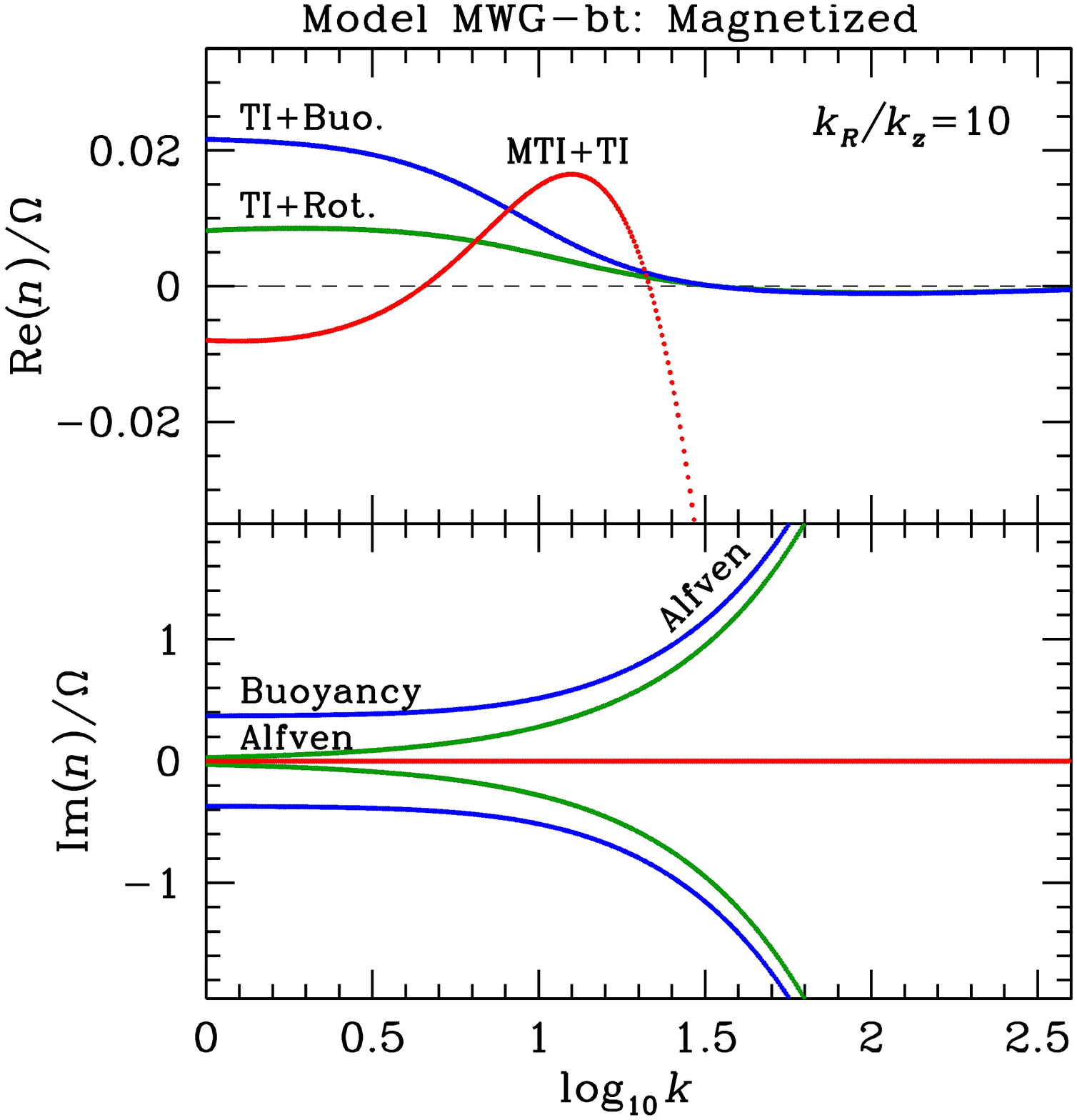,width=0.5\hsize}
\psfig{file=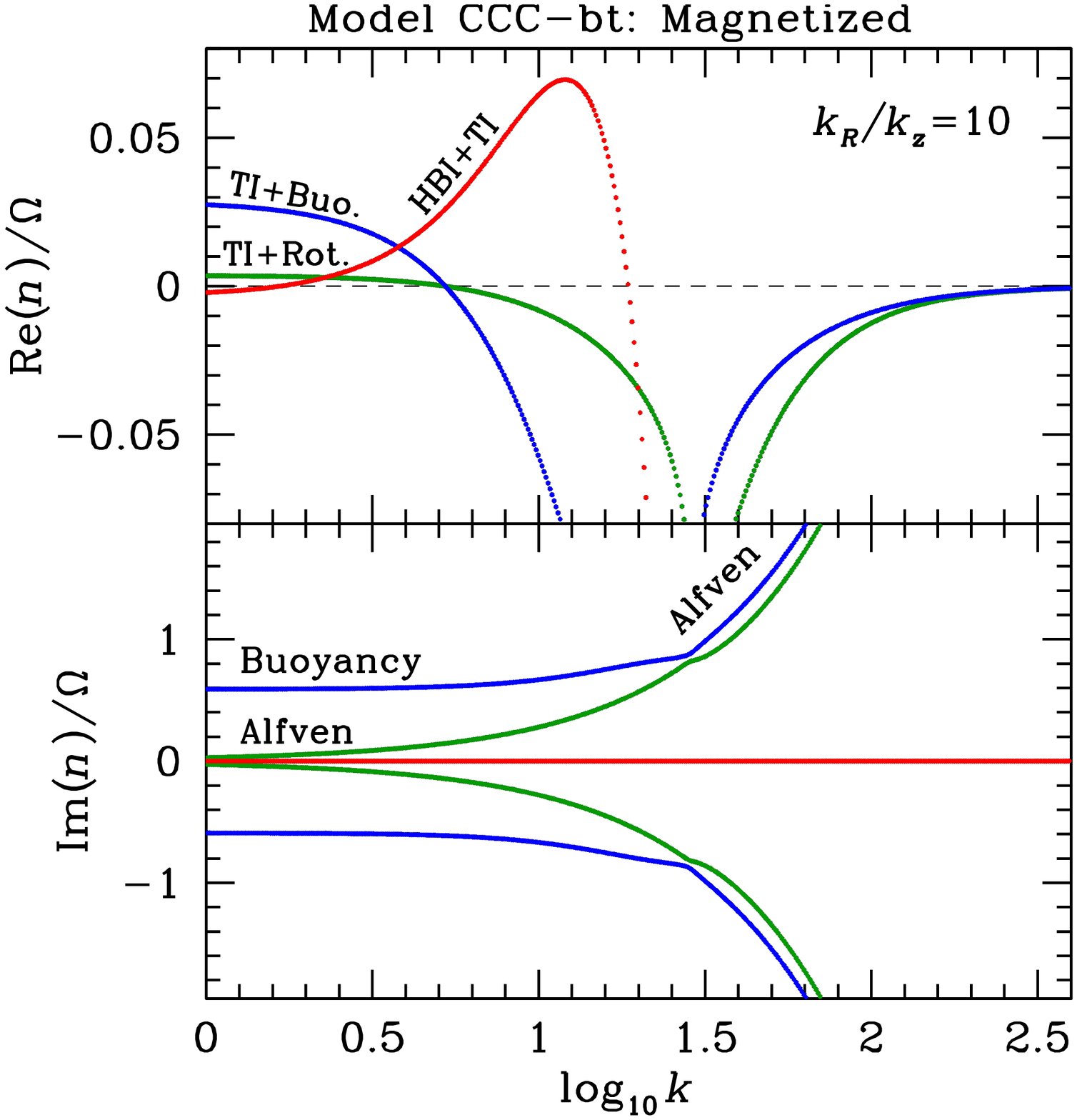,width=0.5\hsize}
}
\caption{Real and imaginary parts of the roots of the dispersion
    relation~(\ref{eq:disp_np13}) as functions of the wave-number $k$
    for models MWG-bt (left-hand panel) and CCC-bt (right-hand panel)
    when $\kR/\kz=10$. Here $k$ is in units of $R^{-1}$. For the sake
    of clarity here we extend the range in $k$ down to $kR=1$, but we
    recall that the results of the linear analysis are rigorous only
    for $kR\gg1$.  Unstable modes are those with ${\rm Re}(n)>0$:
    monotonically growing if ${\rm Im}(n)=0$, over-stable if ${\rm
      Im}(n)\neq0$. We identify the different unstable branches as
    combinations of thermal instability (TI), heat-flux driven buoyancy
    instability (HBI), magneto-thermal instability (MTI), buoyancy and
    rotation. In the imaginary plane we identify the branches in
    terms of Brunt-V\"ais\"al\"a (buoyancy) and Alfv\'en modes. Each
    root is represented with the same color in the upper and lower
    panels.\\}
\label{fig:roots1}
\end{figure}

\subsubsection{Physical nature and growth rates of the unstable modes}
\label{sec:nat}

\begin{figure}
\centerline{
\psfig{file=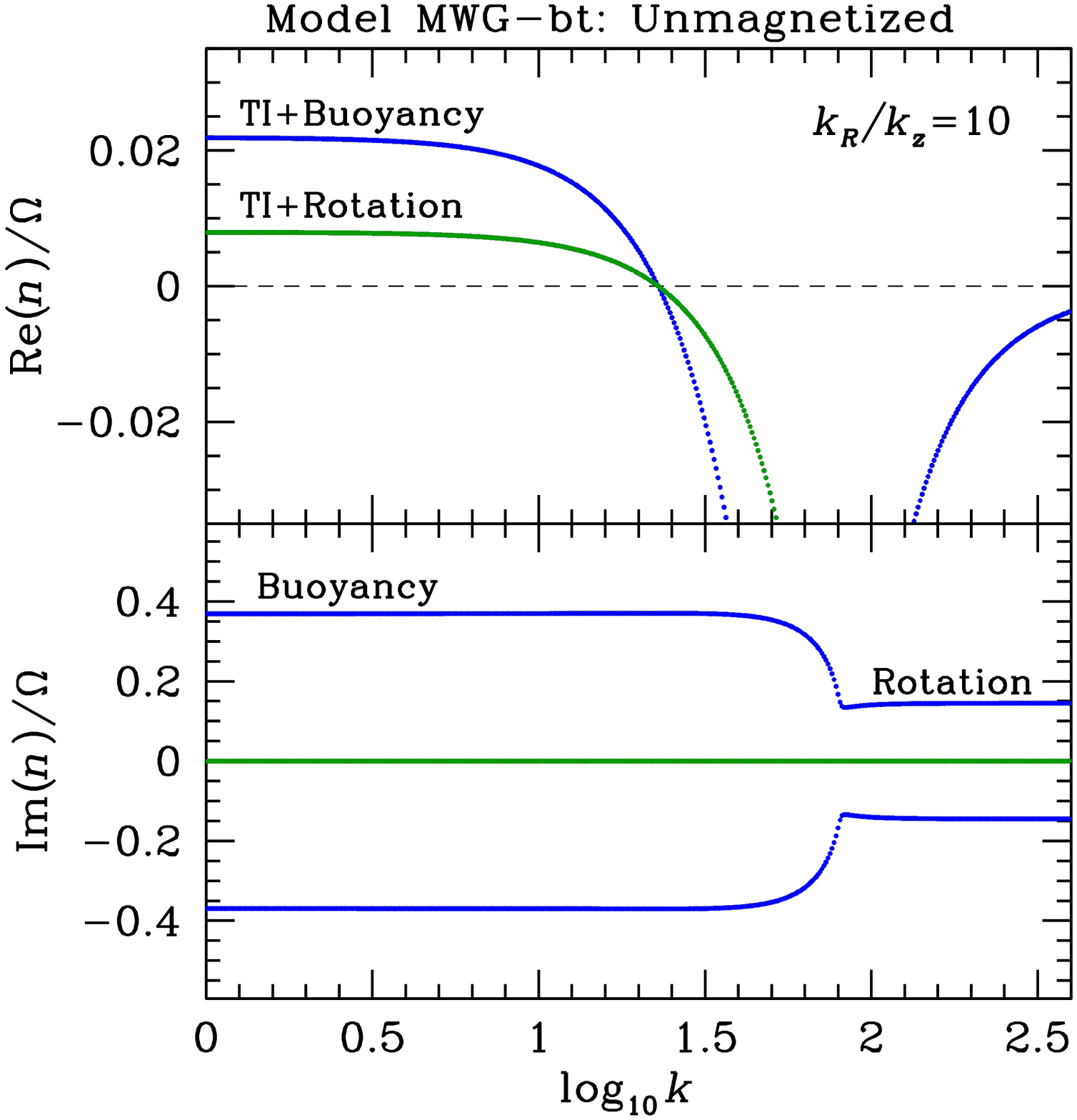,width=0.5\hsize}
\psfig{file=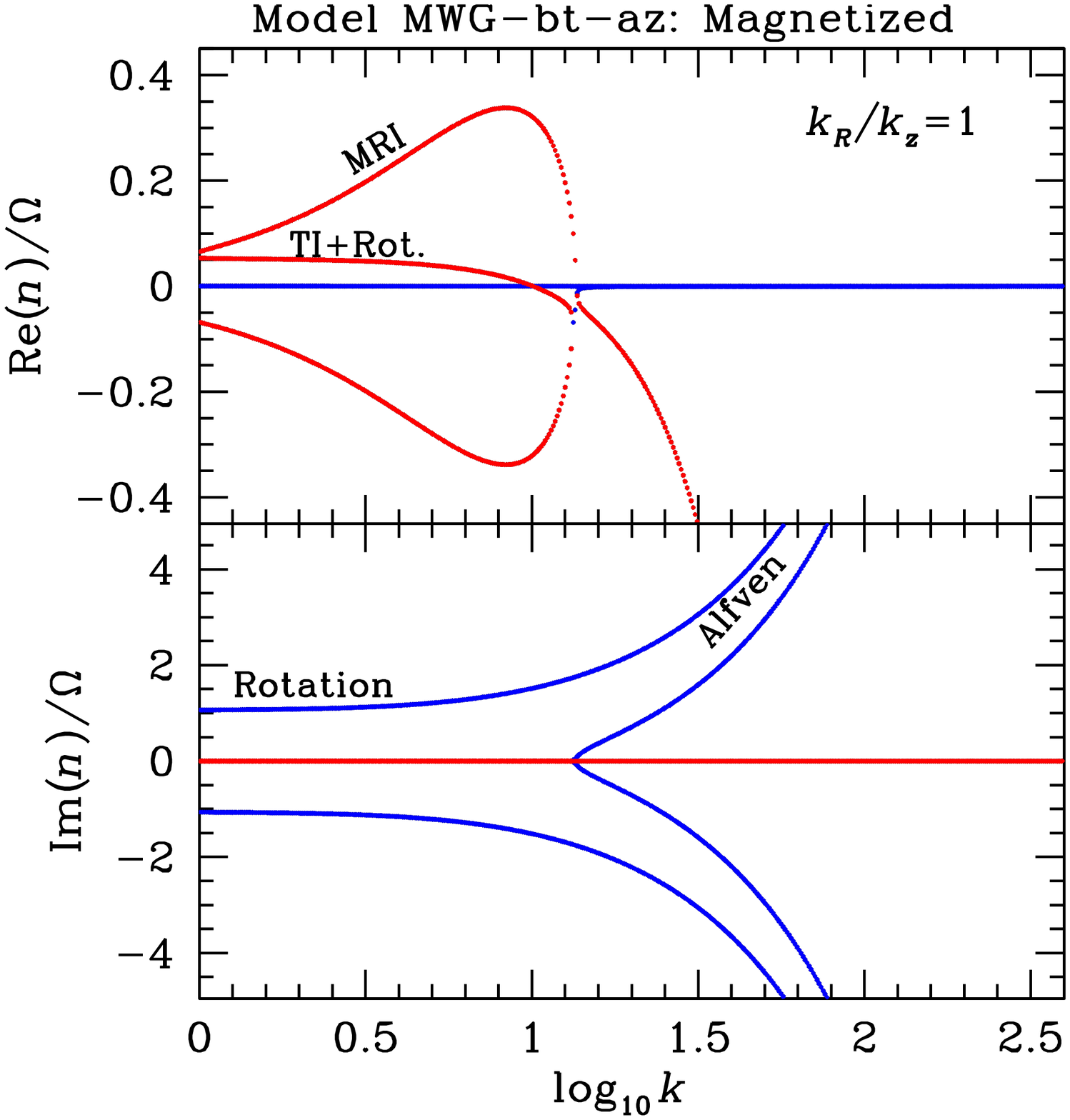,width=0.5\hsize}
}
\caption{Same as Fig.~\ref{fig:roots1}, but, in the left-hand panel,
  for model MWG-bt with no magnetic field and $\kR/\kz=10$ (isotropic
  conduction with $f=0.01$, dispersion relation~\ref{eq:disp_n10}),
  and, in the right-hand panel, for the magnetized model MWG-bt-az and
  $\kR/\kz=1$ (anisotropic conduction, dispersion relation
  \ref{eq:disp_np13}). The branches are identified as in
  Fig.~\ref{fig:roots1}, with the addition of the magnetorotational
  instability (MRI) mode.\\}
\label{fig:roots2}
\end{figure}

\begin{figure}
\centerline{
\psfig{file=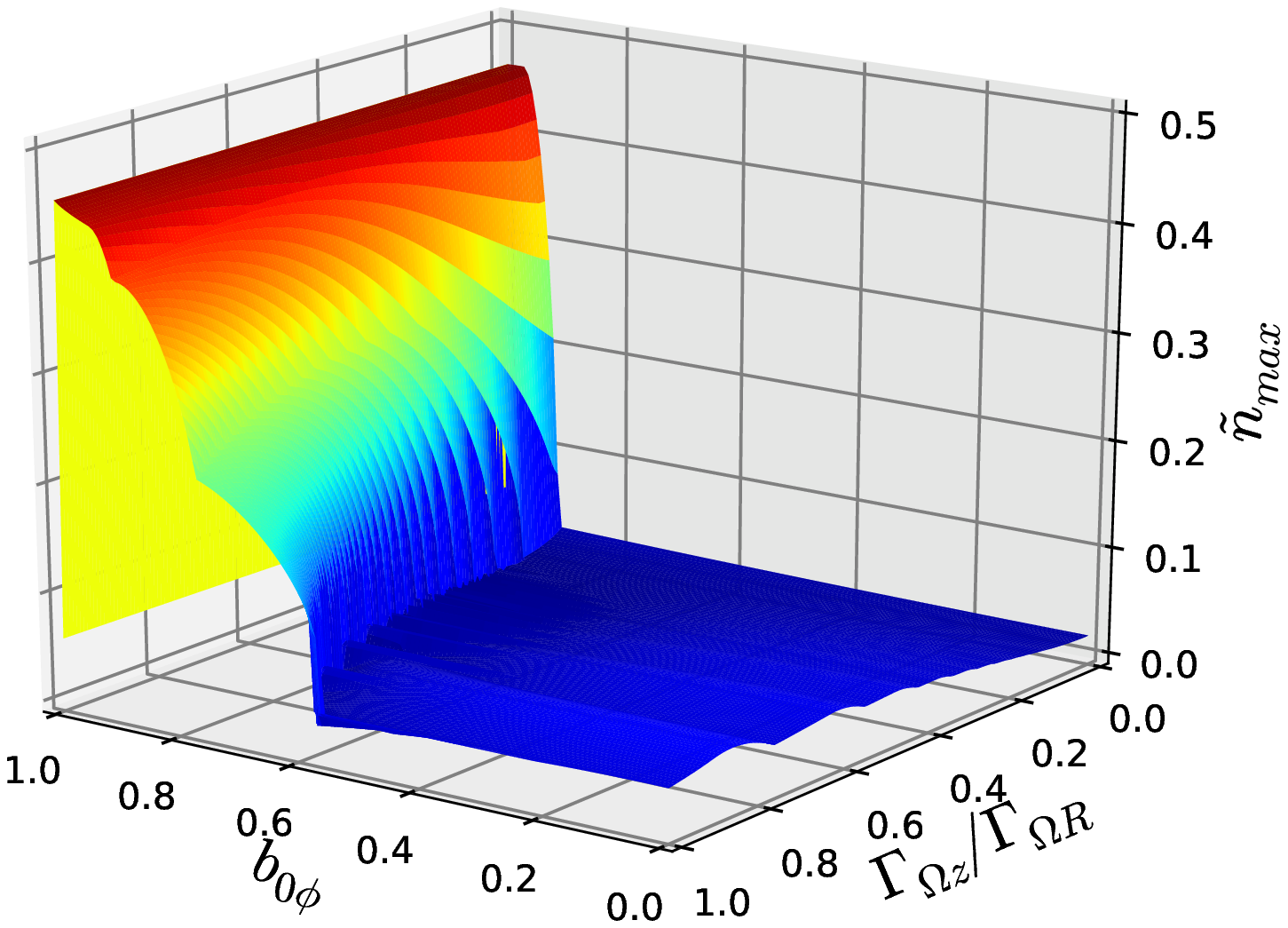,width=0.5\hsize,bbllx=110, bblly=210, bburx=570, bbury=550,clip=}
}
\caption{Maximum growth rate (in units of $\Omega$) of monotonically
  unstable modes as a function of the azimuthal component of the
  magnetic field $\bzerophi$ and of $\GammaOmegaz/\GammaOmegaR$, which
  is a measure of the vertical-to-radial velocity gradient ratio, for
  a set of magnetized Milky-Way galaxy models. The models considered
  here have the same values of the parameters $\GammaOmegaR$,
  $\GammapR$, $\Gammapz$, $\GammaTR$, $\czerotil$, $\omegathtil$,
  $\omegaczerotil$, and $\beta$ as models MWG (see
  Table~\ref{tab:dimlesspar}), but different values of $\GammaOmegaz$
  and $\bzerophi$ (and, as a consequence, different values of the
  dependent parameters $\bzeroR$, $\bzeroz$ and $\GammaTz$). We
  explored wave-vectors with $10\leq k R\leq1000$ and $-25\leq
  \kR/\kz\leq 25$.\\}
\label{fig:rate}
\end{figure}

In order to understand the astrophysical implications of the
instabilities appearing in Figs.~\ref{fig:dom_MWG}-\ref{fig:dom_CCC}
it is important to determine their physical nature. In particular, an
interesting question is whether the instabilities lead to local
condensation and formation of cold clouds (i.e. if they are TIs in the
sense of \citealt{Fie65}). Clearly, the unstable modes here considered
arise from a combination of different physical mechanisms. In fact, as
discussed in \citetalias{Nip13}, even in the absence of radiative
cooling the considered systems are exposed to other instabilities such
as the MRI, the MTI and the HBI. {It is useful to look at the
  behavior of the roots of the dispersion
  relations~(\ref{eq:disp_np13}) and (\ref{eq:disp_n10}) in more
  detail: in particular, here we discuss how the different branches of
  the solutions depend on the wave-number $k$ at fixed radial to
  vertical wave-vector component ratio $x=\kR/\kz$}.  Though several
modes (TI, MTI, HBI, MRI, buoyancy, rotation, Alfv\'en waves) are at
work, for most branches it is possible to identify the dominant modes.
For instance, in Fig.~\ref{fig:roots1} we show the results obtained
for the barotropic models MWG-bt and CCC-bt fixing $x=10$. The
oscillating branches (imaginary parts) are neatly identified with
buoyancy (Brunt-V\"ais\"al\"a) and Alfv\'en modes. Focusing on the
unstable branches, we note that in model MWG-bt (in which temperature
decreases outwards; $\nabla\pzero\cdot\nabla\Tzero>0$) the
monotonically unstable mode is due to a combination of MTI and
TI. However, we verified that the growth rate is dominated by the TI:
the MTI is weaker, because the temperature is relatively low and the
magnetic field lines are non-isothermal (see \citealt{Qua08}), and the
instability is essentially a TI.  In model CCC-bt (in which
temperature increases outwards; $\nabla\pzero\cdot\nabla\Tzero<0$) the
monotonically unstable mode is due to a combination of HBI and TI,
where the HBI is dominant (because of the high temperature).  In both
models there are two over-stable branches (due to a combination of TI
with either rotation or buoyancy), but these modes occur at
sufficiently short wave-lengths ($kR\gtrsim 10$) only for model MWG.
We note that in the over-stable modes the buoyancy frequency is much
higher than the TI frequency, so over-stable modes are not expected to
lead to local condensation \citep{Mal87}. However, the properties of
the monotonically growing modes suggest that local condensation via TI
should be more likely in galactic atmospheres than in cool cores of
galaxy clusters.  A comparison with the roots obtained for model
MWG-bt with no magnetic field (in which only over-stable TI modes are
present and the oscillating branches are buoyancy and rotation modes;
Fig.~\ref{fig:roots2}, left-hand panel) suggests that the presence of
ordered magnetic field tends to promote local condensation through the
combination of the TI with the either the MTI or the HBI.

What is the role of rotation?  In Section~\ref{sec:dom} we anticipated
that in some cases the instability is dominated the MRI. An example is
given in the right-hand panel of Fig.~\ref{fig:roots2}, showing the
roots for model MWG-bt-az fixing $x=1$: it is apparent that the
strongest monotonic instability (maximum growth rate $n/\Omega \approx
0.35$) is not thermal, but MRI. We note that all the plasma models
analyzed in this work are formally MRI unstable because the angular
velocity decreases outwards while the entropy increases outwards (see
\citealt{BalH91}). However, the MRI occurs only for $k$ smaller than a
critical value $\kMRI$ \citep{BalH91}, which in the considered cases
is typically lower than our fiducial limit for short wavelength
$k=10R^{-1}$. In our models only when $\bzerophi$ is close to unity
the MRI occurs at relevant wavelengths, basically because $\kMRI$
increases when the magnetic field components coupled with the
axisymmetric perturbation ($\Bzeroz$ and $\BzeroR$) decrease. We note
that in this case the branch due to the combination of TI and rotation
is monotonically growing, but with substantially smaller growth rate
than the MRI. We expect that these MRI dominated modes do not lead to
local condensation.  Even when the MRI is not dominant, the effect of
rotation on the instabilities studied in this paper is very
important. In the presence of rotation, at least in the simple
configurations here considered in which all the components of the
background magnetic field are time-independent, the magnetic field
lines lie on surfaces of constant angular velocity, which, even in the
absence of cooling, favors the onset of the HBI or of the MTI. In the
presence of cooling the combination of the TI with either the HBI or
the MTI gives rise to monotonically unstable modes, which are expected
to lead to local condensation, provided the plasma temperature is not
too high (otherwise the HBI and the MTI dominate).

The numerical values of the growth rates give additional information
on the astrophysical implications of the studied instabilities. In our
reference range of wave-vector components ($10\leq kR\leq1000$, $-25<x<25$) the
maximum growth rate $\ntilmax\equiv\max\left[{\rm
    Re}(n)\right]/\Omega$ of monotonically growing and over-stable
modes [${\rm Re}(n)>0$] does not exceed $\ntilmax\approx 0.04$ for
models MWG with $\bzerophi=0$.  In the models with azimuthal field
($\bzerophi=0.95$) we get substantially higher growth rates, with
$\ntilmax\approx 0.44$, due to the onset of the MRI.  For models CCC
we find $\ntilmax\approx 0.14$ for monotonic instability (which is
dominated by the HBI) and $\ntilmax\approx 0.03$ for over-stability
(dominated by the TI).  Therefore, in the models here considered, even
for the fastest-growing TI modes the growth-timescale (which is
related to the cooling time) is typically more than one order of
magnitude longer than $\Omega^{-1}$ (which is related to the dynamical
time). Over shorter timescales we expect the HBI to occur in cool
cores and, in specific conditions, the MRI in galactic coronae (and
possibly also in clusters). An overview of the instability growth rate
in a magnetized, rotating Milky-Way like galaxy corona is given in
Fig.~\ref{fig:rate}, where we plot the maximum growth-rate of
monotonically unstable modes $\ntilmax$ as a function of $\bzerophi$
and $\GammaOmegaz/\GammaOmegaR$ for a set of models, which is an
extension of the set MWG. We note that unstable and over-stable modes
are present in all models of this extended set ($0\leq\bzerophi\leq1$
and $0\leq\GammaOmegaz/\GammaOmegaR\leq1$). When $\bzerophi \to 1$ the
growth-rates are maximized (up to $\ntilmax\approx 0.45$), because the
MRI is dominant, but when $\bphi=1$ exactly the growth rate drops by
one order of magnitude, because there is no axisymmetric MRI if
$\bzeroR=\bzeroz=0$.  Figure~\ref{fig:rate} also shows that varying
$\GammaOmegaz/\GammaOmegaR$ has generally a small effect on the
maximum growth rate, though the trend is that $\ntilmax$ is larger for
larger vertical velocity gradients, so baroclinic models tend to be
more unstable than barotropic models (as well known, vertical velocity
gradients are destabilizing even in the absence of magnetic fields;
see, e.g., \citealt{Nel13} and references therein).

For both MWG and CCC models, we have explored the effect of varying
the thermal-to-magnetic pressure ratio $\beta$: the general trend is
that the areas of monotonic instability and over-stability increase
for increasing $\beta$, at the expense of the areas of
stability. However, the effect on the growth rates is small for
variations of $\beta$ up to a factor of a few. Remarkably, if $\beta$
is increased by one order of magnitude or more the MRI tends to be
dominant in all models ($\kMRI\gg R^{-1}$).

\section{Non-axisymmetric perturbations}
\label{sec:nonaxi}

Here we extend the above linear stability analysis to the case of
non-axisymmetric perturbations.  Though formally the existence of
unstable axisymmetric modes is sufficient to infer that the system is
locally unstable, physically it is important to establish whether the
assumption of axisymmetry is a key factor for the onset of the
instability.  For instance, in the unmagnetized case, differentially
rotating systems that are thermally unstable against axisymmetric
perturbations tend to be stable against non-axisymmetric disturbances
(\citetalias{Nip10}).

\subsection{Derivation of the system of ordinary differential equations}
\label{sec:nonaxieq}

The evolution of non-axisymmetric perturbations in a differentially
rotating plasma is more complicated than the axisymmetric case,
because of the effect of the shear (see, e.g.,
\citealt{Cow51,Lin64,Gol65,Ber89,Bal92}).  As in \citetalias{Nip10},
we adopt here shearing coordinates $\phi'=\phi-\Omega(R,z)t$, $R'=R$,
$z'=z$ and $t'=t$ \citep{Gol65,Bal92} to reduce the problem to a
system of ordinary differential equations (ODEs).  In the primed
coordinates the non-axisymmetric perturbations take the form of plane
waves, so we can write the perturbed quantities as
$\Fzero+F(t')\exp(\i\kR' R' + \i\kz' z'+ \i m \phi')$, with $\kR'$,
$\kz'$ and $m$ constant.  {Linearizing
equations~(\ref{eq:mass}-\ref{eq:ene}) in the Boussinesq approximation
(so $T/\Tzero\simeq -\rho/\rhozero$) we get
\begin{eqnarray}
&&\kR\vR+\kz\vz+\kphi\vphi =0,\label{eq:nonaxdiffc}\\
&&
\rhozero{\d \vR\over \d t'}
-2\Omega\rhozero\vphi
+\i\kR p
-\ApR\czero^2\rho
-\frac{\i}{4\pi}(\kv\cdot\Bvzero)\BR
+\frac{\i \kR}{4\pi}(\Bvzero\cdot\Bv)
=0,\\
&&
\rhozero{\d\vphi\over \d t'}
+\left(2\Omega+R'\frac{\de \Omega}{\de R}\right)\rhozero\vR
+R'\frac{\de \Omega}{\de z}\rhozero\vz
+\i \kphi p
-\frac{\i}{4\pi}(\kv\cdot\Bvzero)\Bphi
+\frac{\i \kphi}{4\pi}(\Bvzero\cdot\Bv)
=0,\\
 &&
\rhozero {\d \vz\over \d t'}
+\i\kz p
-\Apz\czero^2\rho
-\frac{\i}{4\pi}(\kv\cdot\Bvzero)\Bz
+\frac{\i \kz}{4\pi}(\Bvzero\cdot\Bv)
=0,\\
&&
 {\d \BR\over \d t'}-\i(\kv\cdot\Bvzero)\vR=0,\\
&&
 {\d \Bphi\over \d t'}-\i(\kv\cdot\Bvzero)\vphi-R'\nabla\Omega\cdot\Bv=0,\\
&&
 {\d \Bz\over \d t'}-\i(\kv\cdot\Bvzero)\vz=0,\\
&&
-{\gamma\over\rhozero}{\d\rho\over \d t'}
+\left(\ApR-\gamma\ArhoR\right)\vR
+\left(\Apz-\gamma\Arhoz\right)\vz
\nonumber\\
&&
\qquad
-\frac{\gamma-1}{\pzero}
 \left[
\frac{\chi(\Tzero) \Tzero(\kv\cdot\bvzero)^2}{\rhozero}
-\left[
\L(\rhozero,\Tzero)+\rhozero\Lrho(\rhozero,\Tzero)-\Tzero\LT(\rhozero,\Tzero)
\right]
\right]\rho
\nonumber\\
&&
\qquad
-\i\frac{(\gamma-1)\chi(\Tzero) \Tzero}{\pzero}
\left[(\kv\cdot\bvzero)\nabla\ln \Tzero
-2(\bvzero\cdot \kv)(\nabla\ln \Tzero \cdot \bvzero)\bvzero
\right]\cdot\bv=0,
\label{eq:nonaxdiffe}
\end{eqnarray}
where we have introduced the quantities
\begin{eqnarray}
&&\ApR\equiv\frac{1}{\pzero}\frac{\pd \pzero}{\pd R},\qquad
\Apz\equiv\frac{1}{\pzero}\frac{\pd \pzero}{\pd z} , \qquad
\ArhoR\equiv\frac{1}{\rhozero}\frac{\pd \rhozero}{\pd R},\qquad
\Arhoz\equiv\frac{1}{\rhozero}\frac{\pd \rhozero}{\pd z},
\end{eqnarray}
and $\kv=\kv(t')\equiv(\kR,\kphi,\kz)$ is the time-dependent
wave-vector, with
\begin{eqnarray}
&&\kR=\kR(t')=\kR'-m t' \frac{\pd \Omega}{\pd R},\\
&&\kz=\kz(t')=\kz'-m t' \frac{\pd \Omega}{\pd z}
\end{eqnarray}
and $\kphi\equiv m/R'$ (independent of time).} In
equation~(\ref{eq:nonaxdiffe}) we have used the divergence-free
condition $\nabla\cdot\Bv=0$ to eliminate the term proportional to
$\kv\cdot\bv$.  The system of
ODEs~(\ref{eq:nonaxdiffc}-\ref{eq:nonaxdiffe}) can be simplified by
eliminating the variables $\vphi$, $p$ and $\Bphi$. In particular, in
the momentum equations, $\vphi$ can be eliminated by using the mass
conservation equation $\vphi=-(\kR\vR+\kz\vz)/\kphi$ and its time
derivative\footnote{The correct expression for ${\d \vphi / \d t'}$ is
  the one reported here and not that given in equation~60 of
  \citetalias{Nip10}, where the last two terms are missing because of
  a typographical error. However, the correct expression for ${\d
    \vphi / \d t'}$ is used in all the calculations of
  \citetalias{Nip10}.}
\begin{eqnarray}
&&{\d \vphi \over \d t'}=-{\kR\over \kphi}{\d \vR \over \d t'}-{\kz\over \kphi}{\d \vz \over \d t'}+R\frac{\de \Omega}{\de R}\vR+R\frac{\de \Omega}{\de z}\vz.
\end{eqnarray}
The pressure perturbation $p$ can be eliminated by combining the
momentum equations as follows: we subtract the $z$ equation
(multiplied by $\kR$) from the $R$ equation (multiplied by $\kz$), and
we subtract the $z$ equation (multiplied by $\kphi$) from the $\phi$
equation (multiplied by $\kz$), so
\begin{eqnarray}
&&
\i\kz p=
-\rhozero {\d \vz\over \d t'}
+\Apz\czero^2\rho
+\frac{\i}{4\pi}(\kv\cdot\Bvzero)\Bz
-\frac{\i \kz}{4\pi}(\Bvzero\cdot\Bv).
\end{eqnarray}
{The divergence-free condition $\nabla\cdot\Bv=0$
(i.e. $\kv\cdot\Bv=0$) can be used to eliminate $\Bphi$, which can be
written as 
\begin{eqnarray}
&&\Bphi=-\frac{\kR}{\kphi}\BR-\frac{\kz}{\kphi}\Bz.
\end{eqnarray}} 
After rearrangement and simplification, we end up with the following
system of five ODEs (in the variables $\vR$, $\vz$, $\rho$, $\BR$ and
$\Bz$):
\begin{eqnarray}
&&{\d \vR \over \d t'}=
{2 \kR \kphi \Omega \over k^2}\left(\GammaOmegaR-{\kz^2\over\kphi^2}\right)\vR+
{2 \kR \kphi \Omega \over k^2}\left(\GammaOmegaz-{\kz\over\kR}{\kz^2+\kphi^2 \over \kphi^2}\right)\vz+
{\czero^2 \ApR\over \rhozero}\left({\kz^2+\kphi^2\over k^2}-{\Apz \over\ApR }{\kR\kz\over k^2}\right)\rho
\nonumber\\
&&
\qquad
+\frac{\i}{4\pi\rhozero}(\kv\cdot\Bvzero)\BR,
\label{eq:system_1}\\
&&{\d \vz \over \d t'}=
{2 \kz \kphi \Omega \over k^2}\left(\GammaOmegaR+{\kR^2+\kphi^2\over\kphi^2}\right)\vR+
{2 \kz \kphi \Omega \over k^2}
\left(
\GammaOmegaz+
{\kz\kR\over\kphi^2}\right)\vz+
{\czero^2 \Apz\over \rhozero}\left({\kR^2+\kphi^2\over k^2}-{\ApR \over\Apz }{\kR\kz\over k^2}\right)\rho
\nonumber\\
&&
\qquad
+\frac{\i}{4\pi\rhozero}(\kv\cdot\Bvzero)\Bz,
\label{eq:system_2}\\
&&
 {\d \BR\over \d t'}-\i(\kv\cdot\Bvzero)\vR=0,
\label{eq:system_3}\\
&&
 {\d \Bz\over \d t'}-\i(\kv\cdot\Bvzero)\vz=0,
\label{eq:system_4}\\
&&
-{\gamma\over\rhozero}{\d\rho\over \d t'}
+\left(\ApR-\gamma\ArhoR\right)\vR
+\left(\Apz-\gamma\Arhoz\right)\vz
\nonumber\\
&&
\qquad
-\frac{\gamma-1}{\pzero}
 \left[
\frac{\chi(\Tzero) \Tzero(\kv\cdot\bvzero)^2}{\rhozero}
-\left[
\L(\rhozero,\Tzero)+\rhozero\Lrho(\rhozero,\Tzero)-\Tzero\LT(\rhozero,\Tzero)
\right]
\right]\rho
\nonumber\\
&&
\qquad
-\i\frac{(\gamma-1)\chi(\Tzero) \Tzero(\kv\cdot\bvzero)}{\pzero}
\left[\nabla\ln \Tzero
-2(\nabla\ln \Tzero \cdot \bvzero)\bvzero
\right]\cdot\bv=0.
\label{eq:system_5}
\end{eqnarray}
Dividing equations~(\ref{eq:system_1}) and (\ref{eq:system_2}) by
$\Omega^2 R$, equations~(\ref{eq:system_3}) and (\ref{eq:system_4}) by
$\Bzero\Omega$, and equation~(\ref{eq:system_5}) by $\gamma\Omega$,
the above system can be rewritten in dimensionless form as
\begin{eqnarray}
&&
{\d \vRtil \over \d \tau}= 
\XRR\vRtil+\XRz\vztil+\XRrho\rhotil+\frac{2\i K\czerotil^2}{\beta}\bR,
\label{eq:nonaxi_adim_1}
\\
&&{\d \vztil \over \d \tau}= 
\XzR\vRtil+\Xzz\vztil+\Xzrho\rhotil+\frac{2\i K\czerotil^2}{\beta}\bz,
\label{eq:nonaxi_adim_2}\\
&&{\d \bR \over \d \tau}=\i K \vRtil,   
\label{eq:nonaxi_adim_3}\\
&&{\d \bz \over \d \tau}=\i K \vztil,   
\label{eq:nonaxi_adim_4}\\
&&
{\d\rhotil\over \d \tau}= 
\XrhoR\vRtil
+\Xrhoz\vztil
+\Xrhorho\rhotil
-\frac{2i K\czerotil^2}{\beta}\YrhoR\bR
-\frac{2i K\czerotil^2}{\beta}\Yrhoz\bz,
\label{eq:nonaxi_adim_5}
\end{eqnarray}
where, for simplicity we have assumed $\bzerophi=0$ and so eliminated
the term proportional to $\bzerophi\bphi$ in the energy equation.  In
the above equations $\vRtil\equiv\vR/\Omega R$,
$\vztil\equiv\vz/\Omega R$, $\rhotil\equiv\rho/\rhozero$, $\tau\equiv
t' \Omega$, $K\equiv (\kvtil\cdot\bvzero)m$, $\kRtil\equiv \kR/\kphi$,
$\kztil\equiv \kz/\kphi$, $\ktil\equiv k/\kphi$, $m=\kphi R'$, and the
quantities indicated with $X$ and $Y$ (defined in
Appendix~\ref{app:list}) are dimensionless functions of $\kvtil(\tau)$
and of the fluid parameters.  {In dimensionless variables we have
\begin{eqnarray}
&&\kRtil(\tau)=\kRtil'-\tau\GammaOmegaR\quad \mbox{and}\quad
  \kztil(\tau)=\kztil'-\tau\GammaOmegaz,
\end{eqnarray}
so ${\d \kRtil}/{\d
  \tau}=-\GammaOmegaR$ and ${\d \kztil}/{\d \tau}=-\GammaOmegaz$.}  It is useful to note that
from the assumed condition of isorotational background field
($\nabla\Omega\cdot\Bvzero=0$) it follows that
$\d\left(\kv\cdot\bvzero\right)/\d\tau=0$, so $K$ is
constant. Differentiating equations~(\ref{eq:nonaxi_adim_1}) and
(\ref{eq:nonaxi_adim_2}) with respect to $\tau$ and using
equations~(\ref{eq:nonaxi_adim_3}) and (\ref{eq:nonaxi_adim_4}) in the
form $\i K \d \bR/\d \tau=-K^2\vR$ and $\i K \d \bz/\d \tau=-K^2\vz$,
we get the following system of ODEs in the variables $\vRtil$,
$\vztil$ and $\rhotil$:
\begin{eqnarray}
&&
{\d^2 \vRtil \over \d \tau^2}= 
(\XRR-\XRrho\YrhoR)\frac{\d \vRtil}{\d \tau}
+(\XRz-\XRrho\Yrhoz)\frac{\d \vztil}{\d \tau}
+\left[
\XRrho\left(\XrhoR+\XRR\YrhoR+\XzR\Yrhoz \right)
+\XRRdot-\frac{2K^2\czerotil^2}{\beta}
\right]\vRtil
\nonumber
\\
&&
\qquad
+\left[
\XRrho\left(\Xrhoz+\XRz\YrhoR+\Xzz\Yrhoz\right)
+\XRzdot
\right]
\vztil
+
\left[
\XRrho\left(\Xrhorho+\XRrho\YrhoR+\Xzrho\Yrhoz\right)
+\XRrhodot\right]
\rhotil,
\label{eq:nonaxi_final_1}
\\
&&{\d^2 \vztil \over \d \tau^2}= 
(\XzR-\Xzrho\YrhoR)\frac{\d \vRtil}{\d \tau}
+(\Xzz-\Xzrho\Yrhoz)\frac{\d \vztil}{\d \tau}
+\left[
\Xzrho\left(\XrhoR+\XRR\YrhoR+\XzR\Yrhoz \right)
+\XzRdot
\right]\vRtil
\nonumber
\\
&&
\qquad
+\left[
\Xzrho\left(\Xrhoz+\XRz\YrhoR+\Xzz\Yrhoz\right)
+\Xzzdot-\frac{2K^2\czerotil^2}{\beta}
\right]
\vztil
+
\left[
\Xzrho\left(\Xrhorho+\XRrho\YrhoR+\Xzrho\Yrhoz\right)
+\Xzrhodot\right]
\rhotil,
\label{eq:nonaxi_final_2}
\\
&&
{\d \rhotil\over \d \tau}=
-\YrhoR\frac{\d \vRtil}{\d \tau}
-\Yrhoz\frac{\d \vztil}{\d \tau}
+\left(\XrhoR+\XRR\YrhoR+\XzR\Yrhoz\right)\vRtil
+\left(\Xrhoz+\XRz\YrhoR+\Xzz\Yrhoz\right)\vztil
\nonumber\\
&&
\qquad
+\left(\Xrhorho+\XRrho\YrhoR+\Xzrho\Yrhoz\right)\rhotil,
\label{eq:nonaxi_final_3}
\end{eqnarray}
where the quantities indicated with $\Xdot$ (reported in
Appendix~\ref{app:list}) are the derivatives with respect to $\tau$ of
the corresponding dimensionless functions indicated with $X$ that we
introduced in
equations~(\ref{eq:nonaxi_adim_1}-\ref{eq:nonaxi_adim_2}).  The final
system of ODEs (\ref{eq:nonaxi_final_1}-\ref{eq:nonaxi_final_3}) in
the variables $\vRtil$, $\vztil$ and $\rhotil$ fully describes the
evolution of non-axisymmetric perturbations.  The system must be
completed by specifying at $\tau=0$ the values of $\vRtil$, $\vztil$,
$\rhotil$, $\d \vRtil/\d \tau$ and $\d\vztil/\d \tau$, which must be
such that equations (\ref{eq:nonaxi_adim_1}-\ref{eq:nonaxi_adim_2})
are satisfied at $\tau=0$.  The coefficients of the system of
equations~(\ref{eq:nonaxi_final_1}-\ref{eq:nonaxi_final_3}) depend on
the following parameters: $\kRtil'$, $\kztil'$, $m$ (characterizing
the perturbation), $\GammaOmegaR$, $\GammaOmegaz$, $\GammapR$,
$\Gammapz$, $\GammaTR$, $\czerotil$, $\omegathtil$, $\omegaczerotil$,
$\beta$ (defining the properties of the background fluid; we recall
that here we are assuming $\bzerophi=0$).  The other parameters
appearing in the coefficients (see Appendix~\ref{app:list}) are not
independent (see Section~\ref{sec:param}).  In particular, we recall
that $\GammarhoR=\GammapR-\GammaTR$ and
$\Gammarhoz=\Gammapz-\GammaTz$.

\begin{figure}
\centerline{
\psfig{file=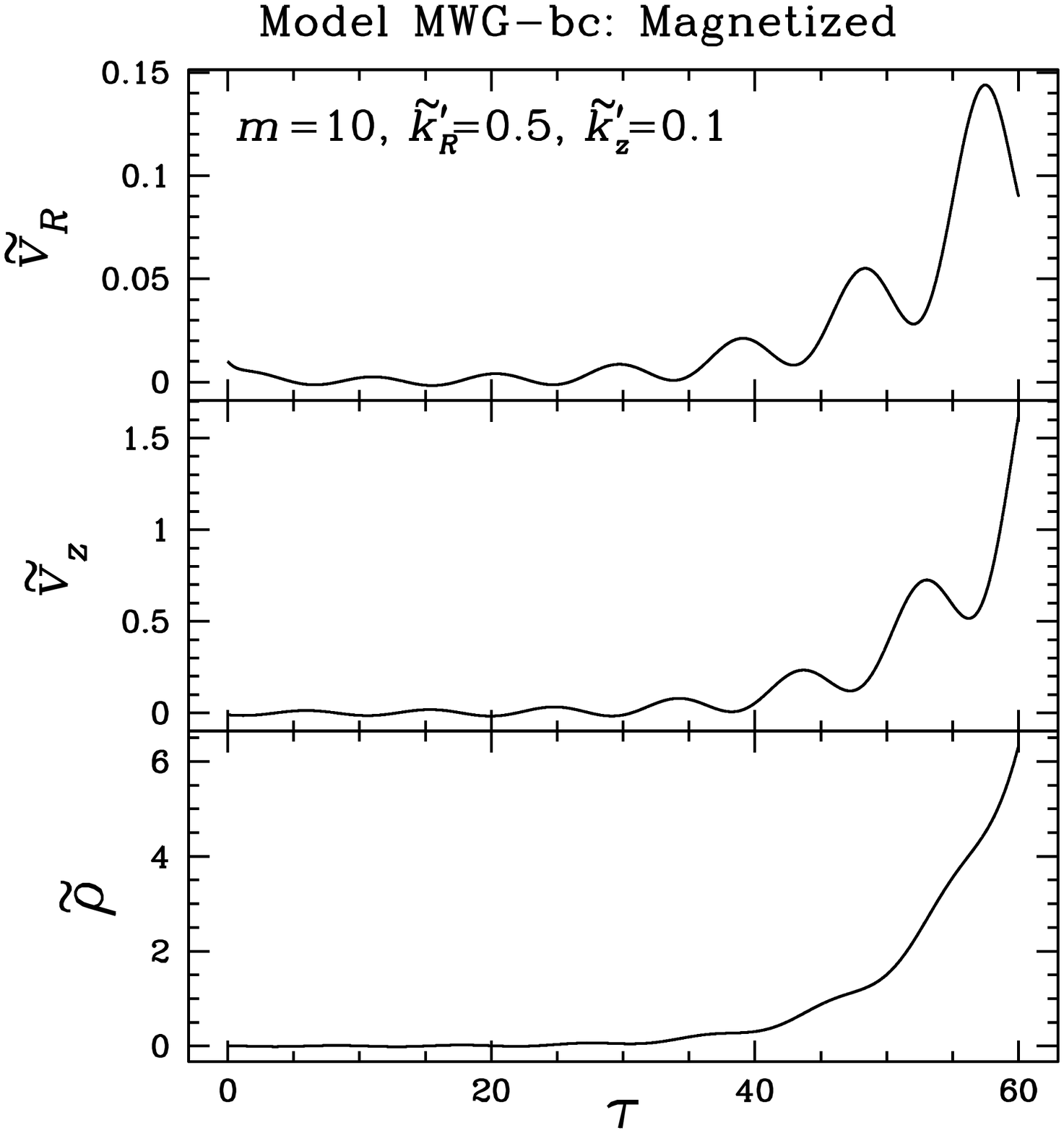,width=0.5\hsize}
\psfig{file=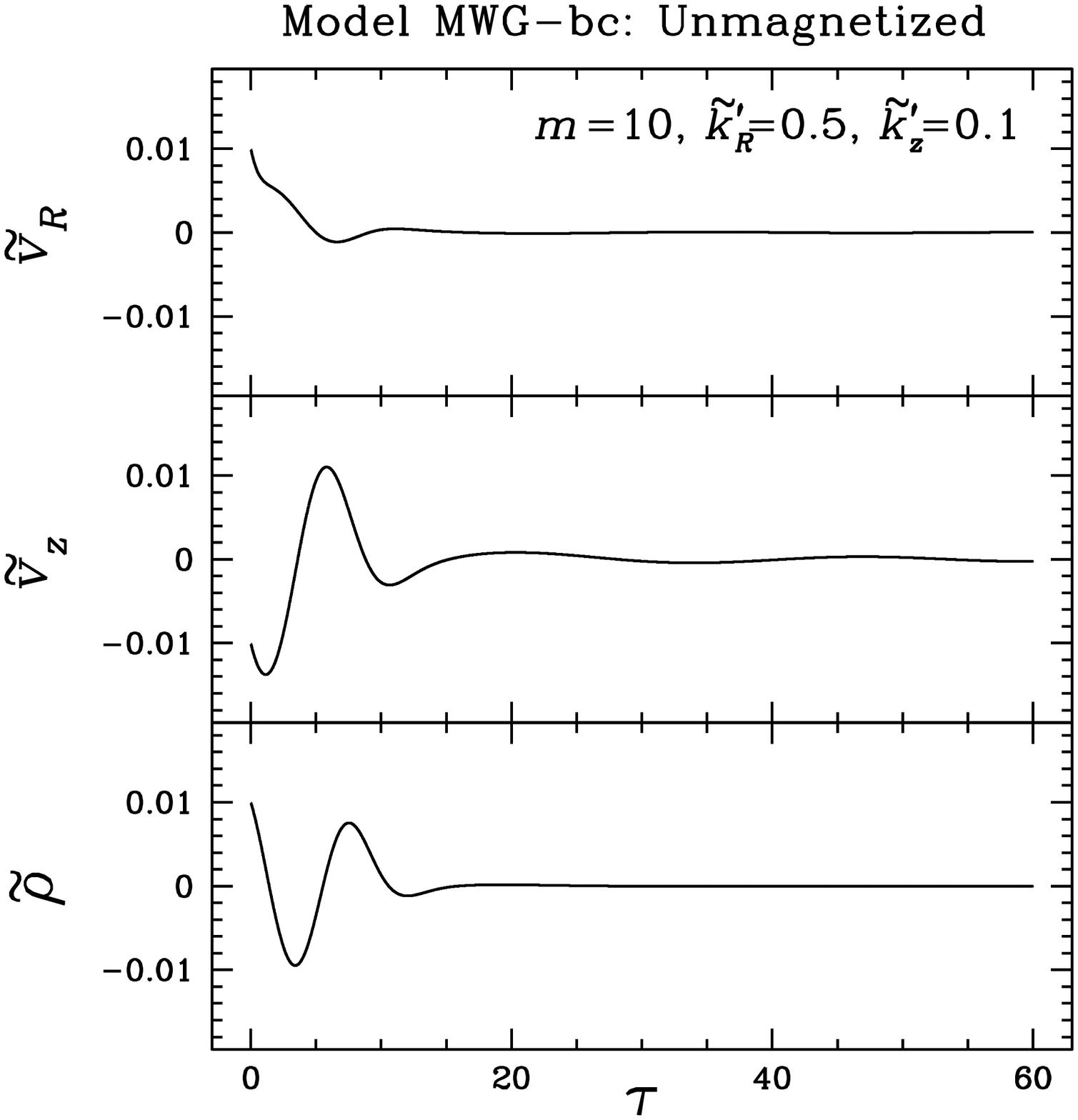,width=0.5\hsize}
}
\caption{Evolution of a non-axisymmetric perturbation for the
  baroclinic Milky-Way like galaxy model MWG-bc in the presence
  (left-hand panel) and in the absence (right-hand panel) of magnetic
  field.  Here $\rhotil$, $\vztil$ and $\vRtil$ are the normalized
  density and velocity perturbations (see Section~\ref{sec:nonaxieq});
  $\tau$ is the time in units of $\Omega^{-1}$. Note the different
  scales in the vertical axes in the two panels. In the unmagnetized
  case we have assumed thermal conduction suppression factor
  $f=0.01$.\\}
\label{fig:ode}
\end{figure}

\begin{figure}
\centerline{ 
  \psfig{file=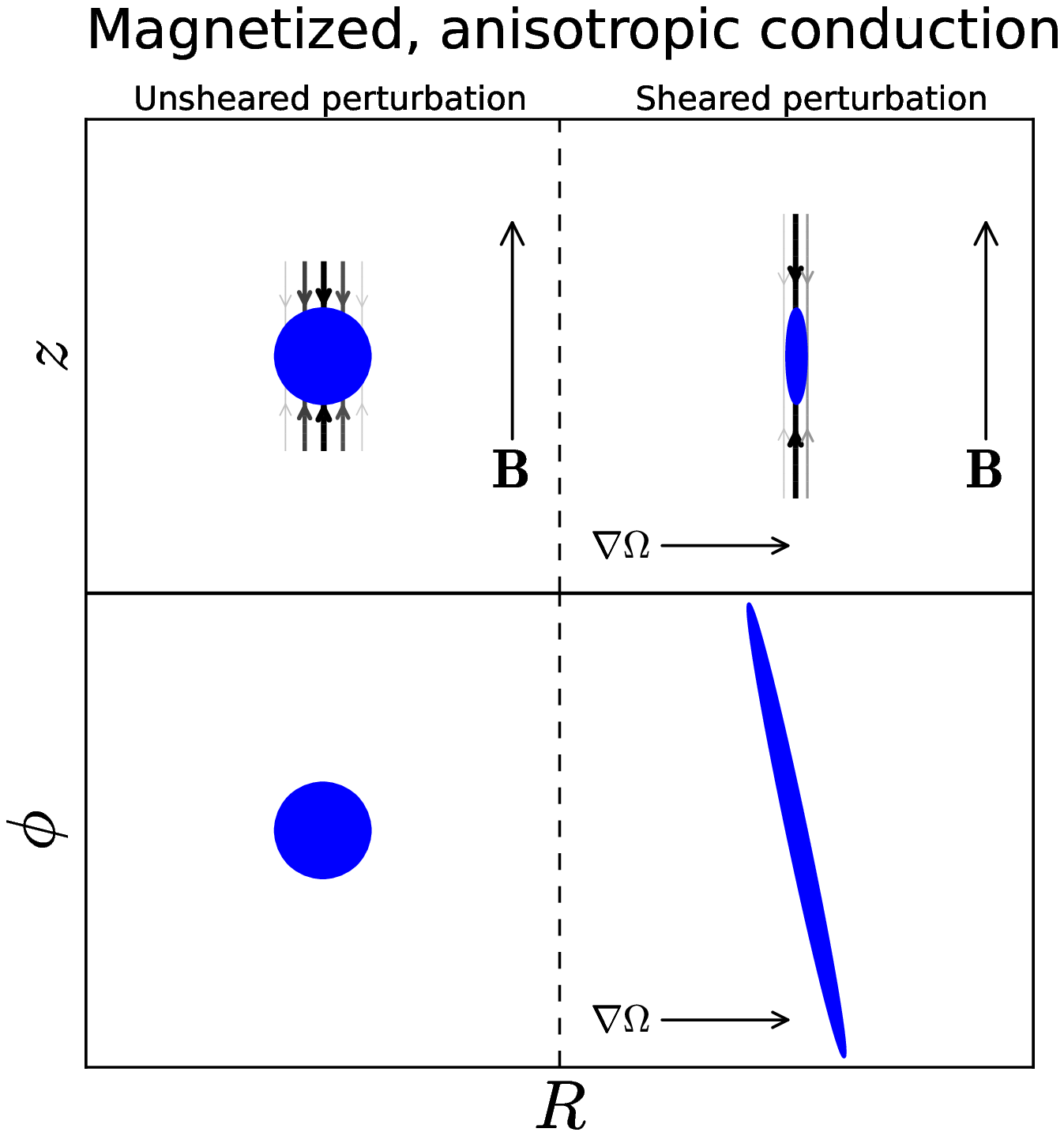,width=0.5\hsize,bbllx=90, bblly=180,
    bburx=530, bbury=613,clip=} 
\psfig{file=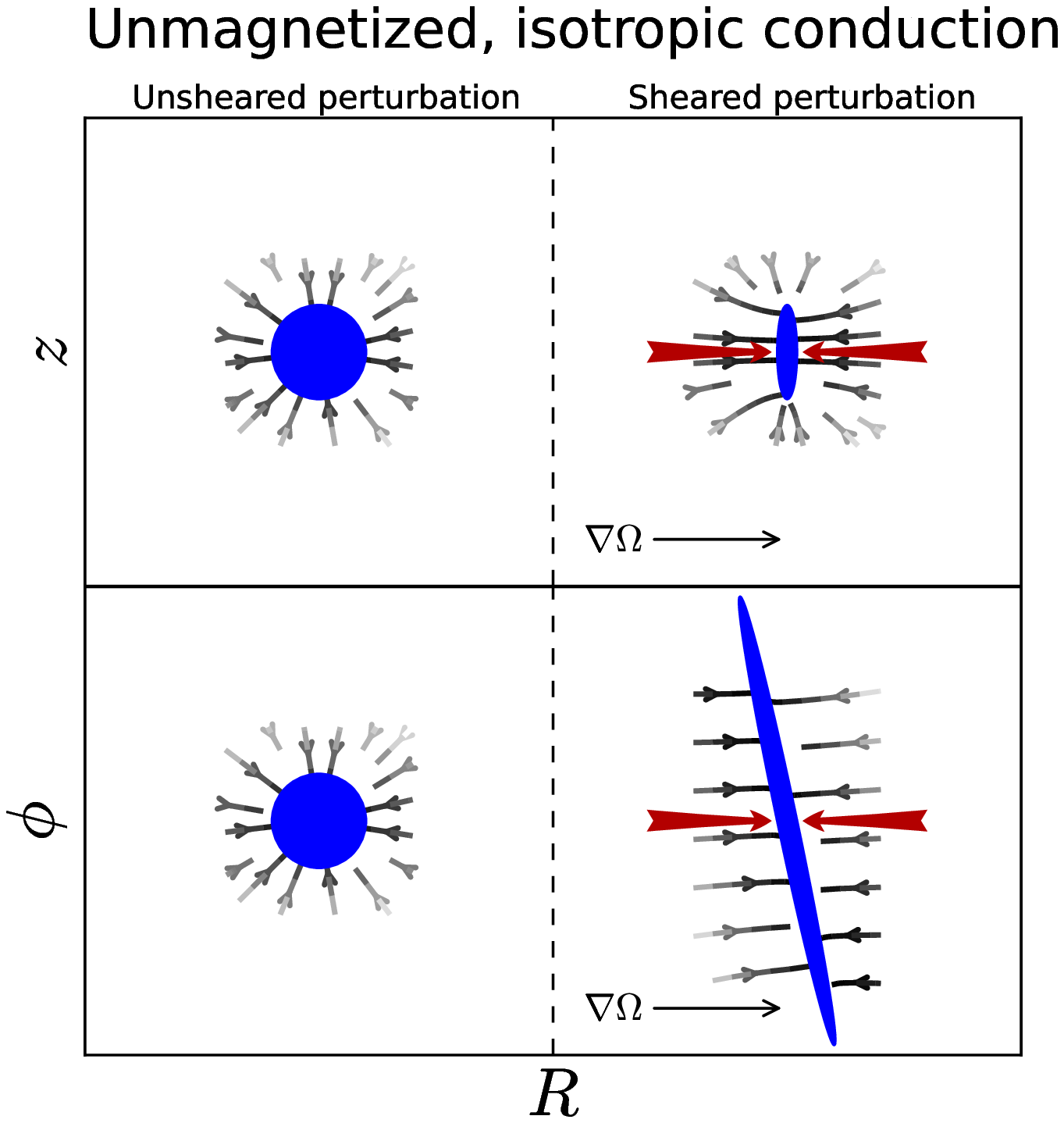,width=0.5\hsize,bbllx=90,
    bblly=180, bburx=530, bbury=613,clip=}}
\caption{ {\it Left-hand panel:} a schematic representation of the
  evolution of a blob-like (non-axisymmetric) thermal disturbance
  (colder than the background) in the presence of differential
  rotation $\Omega=\Omega(R)$ and isorotational magnetic field with no
  azimuthal component. The four quadrants show sections of the initial
  (unsheared; left) and evolved (sheared; right) perturbation in the
  planes $R$-$\phi$ and $R$-$z$. The conductive heat flux $\Qv$ (small
  gray arrows) is anisotropic and it is null across the magnetic field
  lines (i.e. along $R$), so the sheared perturbation is not damped.
  Here we neglect the (small) perturbation in the magnetic field
  direction.  {\it Right-hand panel:} same as the left-hand panel, but
  for an unmagnetized medium.  In this case $\Qv$ is isotropic, so it
  can effectively damp the sheared thermal perturbation, which is
  narrow along $R$ (as indicated by the big red arrows).\\}
\label{fig:cartoon}
\end{figure}

\subsection{Results for non-axisymmetric perturbations}
\label{sec:res_nonaxi}

Given a plasma model, the evolution of a given non-axisymmetric mode
can be calculated by numerically integrating the system of
ODEs~(\ref{eq:nonaxi_final_1}-\ref{eq:nonaxi_final_3}). In particular,
we have solved the system with a fourth-order Runge-Kutta method for
the models MWG-bt, MWG-bc, CCC-bt and CCC-bc, exploring a wide range
of values of wave numbers ($\kRtil'$, $\kztil'$, $m$) and initial
conditions [$\vRtil(0)$, $\vztil(0)$, $\rhotil(0)$, $\d \vRtil/\d \tau
  (0)$, $\d\vztil/\d \tau (0)$]. Overall, the results of these
numerical calculations show that, in the presence of ordered magnetic
field (and therefore anisotropic thermal conduction), the evolution of
non-axisymmetric perturbations with initial wave-vector components
$\kR$, $\kz$ and $k=(\kR^2+\kz^2+\kphi^2)^{1/2}$ is qualitatively
similar to that of axisymmetric perturbations with the same values of
$\kR/\kz$ and $k$. This indicates that the instabilities found in
\citetalias{Nip13} and analyzed in Section~\ref{sec:axi} are not due
to the axisymmetry of the disturbance and confirms that the considered
rotating plasmas are locally unstable to general
perturbations. Remarkably, this is different from the unmagnetized
case, in which non-axisymmetric perturbations are typically stable
even if the corresponding axisymmetric perturbations are unstable
(\citetalias{Nip10}).

The above finding is illustrated in Fig.~\ref{fig:ode}, showing the
evolution of a representative non-axisymmetric perturbation in the
magnetized (anisotropic conduction; left-hand panel) and unmagnetized
(isotropic conduction; right-hand panel) model MWG-bc (for the
unmagnetized system the numerical solution is obtained as described in
\citetalias{Nip10}; equations 64-66 in that paper).  In both cases the
initial wave vector components $\kRtil=0.5$, $\kztil=0.1$, $m=10$
(i.e. $x=5$, $k\simeq 11.2$) correspond to unstable regions for
axisymmetric perturbations (see Section~\ref{sec:dom}) and the initial
conditions are $\vRtil(0)=0.01$, $\vztil(0)=-0.01$ and
$\rhotil(0)=0.01$. In the magnetized model the two additional initial
conditions $\d \vRtil/\d \tau (0)$ and $\d\vztil/\d \tau (0)$ are such
that equations~(\ref{eq:nonaxi_adim_1}-\ref{eq:nonaxi_adim_2}) are
satisfied with $\bv=0$ at $\tau=0$. From Fig.~\ref{fig:ode} it is
apparent that the non-axisymmetric perturbation is unstable (it enters
the non-liner regime at $\tau=t\Omega\simeq45$) when the medium is
magnetized, while it is manifestly stable when the system is
unmagnetized, due to isotropic conduction (even if suppressed by a
factor $f=0.01$).

Physically, the different behavior of non-axisymmetric disturbances
in magnetized and unmagnetized fluids can be understood by considering
the simple case of the evolution of a blob-like (non-axisymmetric)
thermal perturbation in the presence of differential rotation
$\Omega=\Omega(R)$ (see Fig.~\ref{fig:cartoon}). The blob is a small
overdensity close to pressure equilibrium with the hotter surrounding
medium. The effect of the shear is to stretch the perturbation along
the azimuthal direction making it narrow along $R$ (mathematically,
$\kR$ increases with time).  In the magnetized case we have
$\BzeroR=0$ (due to the isorotation condition
$\Bvzero\cdot\nabla\Omega=0$), so, neglecting the small perturbation
in the magnetic field direction, heat conduction does not occur along
$R$ and the perturbation is not damped (left-hand panel in
Fig.~\ref{fig:cartoon}, where for simplicity we assume that the
azimuthal magnetic field component is null). In the absence of
magnetic field thermal conduction is isotropic, so it is effective
along the radial direction and can easily damp the thermal disturbance
which gets thinner because of the shear (right-hand panel in
Fig.~\ref{fig:cartoon}).\\

\section{Summary and conclusions}
\label{sec:con}

In this paper we have studied the nature of local instabilities in
rotating, stratified, weakly magnetized, optically thin astrophysical
plasmas in the presence of radiative cooling and anisotropic thermal
conduction. A summary of the main results of the present work is the
following.
\begin{enumerate}
\item We have provided the equations that allow to determine the
  linear evolution of axisymmetric and non-axisymmetric perturbations
  at any position of a differentially rotating plasma in the presence
  of a weak ordered magnetic field. Given a model for the background
  plasma, the evolution of axisymmetric perturbations can be computed
  by solving numerically the dispersion
  relation~(\ref{eq:disp_np13}). The evolution of non-axisymmetric
  perturbations can be determined by integrating numerically the
  system of ODEs~(\ref{eq:nonaxi_final_1}-\ref{eq:nonaxi_final_3}).
\item 
 We have studied the stability properties of rotating models
 representative of cool cores of galaxy clusters and coronae of Milky
 Way-like galaxies.  In all cases we found monotonically unstable
 axisymmetric modes. The instability is dominated by the HBI in cool
 cores and by a combination of TI and MTI in galactic coronae (with
 the exception of models with very weak poloidal component of the
 magnetic field, in which the MRI is the dominant instability).
\item For the same galaxy and galaxy-cluster models, we have computed
  the linear evolution of several non-axisymmetric disturbances,
  finding that the linear non-axisymmetric modes behave similarly to
  axisymmetric modes with the same wave-length and orientation in the
  meridional plane, so these systems are locally unstable against
  general perturbations. In particular, in contrast with the
  unmagnetized case, differential rotation does not stabilize
  blob-like disturbances in the presence of an ordered magnetic field.
\item Overall, magnetized systems are more prone to local TI than
  unmagnetized systems: in particular, thermal perturbations tend to
  be effectively damped by isotropic heat conduction in unmagnetized
  systems, while, under certain conditions, they can grow
  monotonically and lead to local condensations in the presence of
  magnetic fields.
\item Differential rotation plays a crucial role in the studied
  instabilities. Remarkably, if the magnetic field is sufficiently
  weak the MRI is dominant even in pressure-supported systems such as
  galactic coronae and cool cores of galaxy clusters.  But also when
  the MRI is not strong, differential rotation can favor the onset of
  either the MTI or the HBI, which, combined with the TI, can lead to
  local condensation of cold gas.  The presence of vertical velocity
  gradient is destabilizing, so baroclinic models tend to be more
  unstable than barotropic models.
\end{enumerate}

The original motivation of the present investigation was the question
of whether cold clouds can form spontaneously throughout the
virial-temperature atmospheres of galaxies and clusters of
galaxies. In this work we have focused on differentially rotating
plasmas, because rotation is potentially relevant in the atmospheres
of both galaxies \citep{Mar11} and galaxy clusters \citep{Bia13}.  A
necessary condition for the spontaneous formation of cold clouds via
TI is that, at least in the linear regime, thermal perturbations grow
monotonically.  Our results suggest that in the presence of a weak
ordered magnetic field, provided that the MRI is not dominant and that
the gas temperature is relatively low, thermal perturbations can lead
to local condensation through a combination of the TI with either the
MTI or the HBI. Therefore the formation of cold clouds via local TI is
hampered in the cluster cool cores, while it is possible under
specific conditions in galactic coronae. While the gas temperature of
galactic and galaxy-cluster atmospheres is relatively well constrained
(either observationally or theoretically), much less is known about
the distribution of their specific angular momentum and the geometry
of their magnetic field. In the hypothesis that the conditions are
such that linear thermal perturbations grow monotonically, we are left
with the question of the non-linear evolution of these unstable
systems, which could be addressed with MHD simulations (see, for the
non-rotating case, \citealt{Kun12}, \citealt{Mcc12} and
\citealt{Wag14}). One possibility is that, in the non-linear regime,
finite-size cold clouds form and the medium becomes multiphase, but it
is also possible that the main outcome of the instability is that the
magnetic field is rearranged in a configuration that counteracts the
development of further instabilities or that the gas becomes highly
turbulent and the magnetic field highly tangled.

A limitation of the present work is that, even in the presence of
ordered magnetic fields, we have assumed for simplicity pressure to be
isotropic, neglecting the \citet{Bra65} viscosity.  In fact, in a
magnetized plasma anisotropic momentum transport can affect the
stability properties of the plasma: for instance, studying
non-rotating models of cluster cool cores, \citet{Lat12} and
\citet{Kun12} concluded that, in the presence of \citet{Bra65}
viscosity, the HBI is substantially reduced, being localized only in
the inner 20\% of the cluster core. The question of whether and how
much the results of the present work are affected by \citet{Bra65}
viscosity can be addressed by extending the calculations to the case
in which anisotropic pressure is self-consistently included: such an
investigation would represent a natural follow-up of this paper.

\acknowledgments

We are grateful to Steven Balbus for useful discussions.  We
acknowledge financial support from PRIN MIUR 2010-2011, project ``The
Chemical and Dynamical Evolution of the Milky Way and Local Group
Galaxies'', prot. 2010LY5N2T.

\appendix

\section{List of definitions}
\label{app:list}

Here is a list of definitions of quantities used in Section \ref{sec:nonaxi}:
\begin{eqnarray}
&&
\XRR\equiv{2 \kRtil \over \ktil^2}\left(\GammaOmegaR-\kztil^2\right),
\nonumber\\
&&
\XRz\equiv{2 \kRtil\over \ktil^2}\left[\GammaOmegaz-{\kztil\over\kRtil}(\kztil^2+1)\right],
\nonumber\\
&&
\XRrho\equiv
{\czerotil^2 \GammapR}\left({\kztil^2+1\over \ktil^2}-{\Gammapz \over\GammapR }{\kRtil\kztil\over \ktil^2}\right),
\nonumber\\
&&
\XzR\equiv{2 \kztil  \over \ktil^2}\left(\GammaOmegaR+\kRtil^2+1\right),
\nonumber\\
&&
\Xzz\equiv{2 \kztil\over \ktil^2}
\left(
\GammaOmegaz+
\kztil\kRtil\right),
\nonumber\\
&&
\Xzrho\equiv{\czerotil^2 \Gammapz}\left({\kRtil^2+1\over \ktil^2}-{\GammapR \over\Gammapz }{\kRtil\kztil\over \ktil^2}\right),
\nonumber\\
&&
\XrhoR\equiv\frac{1}{\gamma}\left(\GammapR-\gamma\GammarhoR\right),
\nonumber\\
&&
\Xrhoz\equiv\frac{1}{\gamma}\left(\Gammapz-\gamma\Gammarhoz\right),
\nonumber\\
&&
\Xrhorho\equiv- \left( \omegaczerotil K^2 +\omegathtil \right),
\nonumber\\
&&
\YrhoR\equiv\frac{\beta\omegaczerotil}{2\czerotil^2}\left[\GammaTR-2(\GammaTR\bzeroR+\GammaTz\bzeroz)\bzeroR\right],
\nonumber\\
&&
\Yrhoz\equiv\frac{\beta\omegaczerotil}{2\czerotil^2} \left[\GammaTz-2(\GammaTR\bzeroR+\GammaTz\bzeroz)\bzeroz\right],
\nonumber\\
&&
\XRRdot\equiv\frac{\d \XRR}{\d \tau}=
\frac{2}{\ktil^4}
\Big[
\GammaOmegaR\kztil^4
-\GammaOmegaR\left(\kRtil^2+\GammaOmegaR-1\right)\kztil^2
+2\GammaOmegaz(\kRtil+\GammaOmegaR+1)\kRtil\kztil
+\GammaOmegaR^2(\kR^2-1)
\Big],
\nonumber\\
&&
\XRzdot\equiv\frac{\d \XRz}{\d \tau}=
\frac{2}{\ktil^4}
\Big[
\GammaOmegaz\kztil^4
-2\GammaOmegaR\kRtil\kztil^3
+\GammaOmegaz(3\kRtil^2+2-\GammaOmegaR)\kztil^2
\nonumber\\
&&\qquad
+2\left(\GammaOmegaz^2-\GammaOmegaR\right)\kRtil\kztil
+\GammaOmegaz\left(\GammaOmegaR+1\right)\kRtil^2+\GammaOmegaz\left(1-\GammaOmegaR\right)
\Big],
\nonumber\\
&&
\XRrhodot\equiv\frac{\d \XRrho}{\d \tau}=
\frac{\czerotil^2\GammapR}{\ktil^4}
\Bigg[
2\kRtil
\left(
\GammaOmegaR\kztil^2-\GammaOmegaz\kRtil\kztil+\GammaOmegaR
\right)
\nonumber\\
&&\qquad
+\frac{\Gammapz}{\GammapR}
\Big[
\GammaOmegaR\kztil^3-\GammaOmegaz\kRtil\kztil^2
+(1-\kRtil^2)\GammaOmegaR\kztil
+\GammaOmegaz\kRtil(\kRtil^2+1)
\Big]
\Bigg],
\nonumber\\
&&
\XzRdot\equiv\frac{\d \XzR}{\d \tau}=
\frac{2}{\ktil^4}
\Big[
-2\GammaOmegaR\kRtil\kztil^3
+\GammaOmegaz\left(\kRtil^2+\GammaOmegaR+1\right)\kztil^2
+2\GammaOmegaR^2\kRtil\kztil
-\GammaOmegaz(\kRtil^2+1)\left(\kRtil^2+\GammaOmegaR+1\right)
\Big],
\nonumber\\
&&
\Xzzdot\equiv\frac{\d \Xzz}{\d \tau}=
\frac{2}{\ktil^4}
\Big[
-\GammaOmegaR\kztil^4
+\left(\GammaOmegaR\kRtil^2+\GammaOmegaz^2-\GammaOmegaR\right)\kztil^2
-2\GammaOmegaz\kRtil\left(\kRtil^2-\GammaOmegaR+1\right)\kztil
-\GammaOmegaz^2\left(\kRtil^2+1\right)
\Big],
\nonumber\\
&&
\Xzrhodot\equiv\frac{\d \Xzrho}{\d \tau}=
\frac{\czerotil^2\Gammapz}{\ktil^4}
\Bigg[
-2\kztil
\left(
\GammaOmegaR\kztil\kRtil-\GammaOmegaz\kRtil^2-\GammaOmegaz
\right)
\nonumber\\
&&\qquad
+\frac{\GammapR}{\Gammapz}
\Big[
\GammaOmegaR\kztil^3-\GammaOmegaz\kRtil\kztil^2
+(1-\kRtil^2)\GammaOmegaR\kztil
+\GammaOmegaz\kRtil(\kRtil^2+1)
\Big]
\Bigg].
\nonumber
\end{eqnarray}

\end{document}